\documentclass[authoryear,preprint,review,10pt,3p]{elsarticle}

\usepackage{amssymb}
\usepackage{amsmath}
\usepackage{amsfonts}
\usepackage{latexsym}
\usepackage{bm}
\usepackage{ascmac}
\usepackage{subfigure}
\usepackage{natbib}
\usepackage{lscape}
\usepackage{url}
\usepackage{flushend}
\usepackage[footnotesize]{caption}
\usepackage{booktabs}
\usepackage[colorlinks=true,urlcolor=black,bookmarks=false,bookmarksnumbered=false,bookmarksopen=false,
pdfborder={0 0 0},bookmarkstype=toc,linkcolor=black,citecolor=black,linktocpage=true,draft=false]{hyperref}

\newcommand{\vm}[1]{\mbox{\textbf{#1}}}

\newcommand{\te}{\mathrm{T}}

\journal{NeuroImage}

\begin{document}

\begin{frontmatter}



\title{Fluctuations between high- and low-modularity topology in time-resolved functional connectivity}

\author{$^{\mathrm{a}}$Makoto Fukushima$^*$}
\author{$^{\mathrm{a}, \mathrm{b}}$Richard F. Betzel}
\author{$^{\mathrm{a}, \mathrm{c}}$Ye He}
\author{$^{\mathrm{d}}$Marcel A. de Reus}
\author{$^{\mathrm{d}}$Martijn P. van den Heuvel}
\author{$^{\mathrm{c}}$Xi-Nian Zuo}
\author{$^{\mathrm{a}, \mathrm{e}}$Olaf Sporns}
\address{$^{\mathrm{a}}$Department of Psychological and Brain Sciences, Indiana University, Bloomington, IN, 47405, USA}
\address{$^{\mathrm{b}}$Department of Bioengineering, University of Pennsylvania, Philadelphia, PA, 19104, USA}
\address{$^{\mathrm{c}}$CAS Key Laboratory of Behavioral Science, Institute of Psychology, Beijing 100101, China}
\address{$^{\mathrm{d}}$Department of Psychiatry, Brain Center Rudolf Magnus, University Medical Center Utrecht, Utrecht, the Netherlands}
\address{$^{\mathrm{e}}$Indiana University Network Science Institute, Bloomington, IN, 47405, USA\vspace{-4.5mm}}
\cortext[cor1]{Corresponding author at: Department of Psychological and Brain Sciences, Indiana University, Bloomington, IN, 47405, USA.
{\it Email address:} {\tt mfukushi@indiana.edu} (Makoto Fukushima)}

\begin{abstract}
Modularity is an important topological attribute for functional brain networks.
Recent human fMRI studies have reported that modularity of functional networks varies not only 
across individuals being related to demographics and cognitive performance, but also within 
individuals co-occurring with fluctuations in network properties of functional connectivity, 
estimated over short time intervals.
However, characteristics of these time-resolved functional networks during periods of high and low 
modularity have remained largely unexplored.
In this study we investigate basic spatiotemporal properties of time-resolved networks in the high 
and low modularity periods during rest, with a particular focus on their spatial connectivity 
patterns, temporal homogeneity and test-retest reliability.
We show that spatial connectivity patterns of time-resolved networks in the high and low modularity 
periods are represented by increased and decreased dissociation of the default mode network module 
from task-positive network modules, respectively.
We also find that the instances of time-resolved functional connectivity sampled from within the 
high (respectively, low) modularity period are relatively homogeneous (respectively, heterogeneous) 
over time, indicating that during the low modularity period the default mode network interacts 
with other networks in a variable manner.
We confirmed that the occurrence of the high and low modularity periods varies across individuals 
with moderate inter-session test-retest reliability and that it is correlated with 
previously-reported individual differences in the modularity of functional connectivity estimated 
over longer timescales.
Our findings illustrate how time-resolved functional networks are spatiotemporally organized 
during periods of high and low modularity, allowing one to trace individual differences in 
long-timescale modularity to the variable occurrence of network configurations at shorter timescales.
\end{abstract}


\end{frontmatter}


\section{Introduction}
Spontaneous brain activity has been shown to form highly organized network patterns of global 
functional couplings \citep{Biswal1995,Fox2007}.
These so-called resting-state functional brain networks have been consistently identified from 
fluctuations in the blood oxygenation level dependent (BOLD) signal measured over several minutes of 
resting-state functional magnetic resonance imaging (rs-fMRI) 
\citep{Damoiseaux2006,Fox2005,Smith2009,VanDijk2010} as well as from spontaneous 
magnetoencephalographic recordings \citep{Brookes2011,Hipp2012}.
The coupling strengths between brain regions in these functional networks (i.e., functional 
connectivity) are typically defined based on statistical dependencies (e.g., correlation coefficient) 
of those regional activities.
Functional connectivity as measured by rs-fMRI is known to have functional significance and clinical 
relevance \citep{Greicius2004,Seeley2007,Buckner2009}.

Many topological properties of brain networks have been identified using graph theoretical analysis 
\citep{Bullmore2009,Rubinov2010}.
In particular, there is strong evidence for modules both in functional and structural brain networks 
(\citealp{Hagmann2008,Meunier2009b,Heuvel2011,Power2011,Shen2012,Heuvel2013}; for a review, see 
\citealp{Sporns2016}).
The modules are defined as subnetworks of densely connected or coupled nodes that are only sparsely 
connected or coupled with the rest of the network.
In general, modular organization of biological networks may confer several functional advantages, 
e.g., adaptability to a changing environment \citep{Kashtan2005,Kashtan2007} and conservation of 
resources for connections or couplings \citep{Bullmore2012,Clune2013}.
Modularity refers to the degree to which modules dissociate from each other and is a particularly 
important topological attribute for functional brain networks.
Previous studies have reported that modularity of functional networks in the human brain varies 
across individuals and is related to life-span development and aging 
(\citealp{Betzel2014,Cao2014b,Chan2014,Geerligs2015,Gallen2016}; for a review, see \citealp{Zuo2017}), 
and cognitive performance, e.g., in working memory 
\citep{Kitzbichler2011,Stevens2012,Meunier2014,Stanley2014,Vatansever2015}.
Modules in functional networks have also been shown to reconfigure with changes in cognitive and 
behavioral states \citep{Bassett2011,Braun2015}.

Recently, functional connectivity in rs-fMRI data has been reported to fluctuate on a time scale of 
tens of seconds (\citealp{Chang2010,Handwerker2012,Hutchison2013a,Keilholz2013,Gonzalez-Castillo2014}; 
for reviews, see \citealp{Hutchison2013}, \citealp{Calhoun2014}, and \citealp{Preti2016}).
Fluctuations on this short timescale or {\it time-resolved} functional connectivity have been 
observed not only at the level of individual edges, but also at the level of the whole network. 
Network-level fluctuations have been assessed by focusing on temporal changes in various types 
of graph metrics \citep{Jones2012,Zalesky2014,Betzel2016,Chen2016,Shine2016,Shine2016a} and by 
decomposing the time series of time-resolved functional connectivity into a small set of connectivity 
states or components \citep{Leonardi2013,Allen2014,Damaraju2014,Leonardi2014,Yang2014,Barttfeld2015,Hutchison2015,Miller2016,Rashid2016,Nomi2016}.
Fluctuations in time-resolved functional connectivity has been shown to be related to development 
\citep{Hutchison2015}, consciousness \citep{Barttfeld2015}, cognition and behavior 
\citep{Kucyi2014,Yang2014,Madhyastha2015,Chen2016,Shine2016,Shine2016a,Nomi2016}, and 
neuropsychiatric disorders \citep{Damaraju2014,Miller2016,Rashid2016,Zhang2016}.
Changes in functional networks during rest have been observed not only from relatively slow 
BOLD signals but also from relatively fast electromagnetic recordings 
\citep{DePasquale2010,DePasquale2012,DePasquale2016,Baker2014,Brookes2014}.

In addition to previously observed individual variations in modularity measured over longer 
timescales, several recent human rs-fMRI studies have reported that modularity of time-resolved 
functional networks varies on shorter timescales \citep{Jones2012,Betzel2016,Shine2016}.
\cite{Jones2012} applied a community detection algorithm to time-resolved networks and demonstrated 
temporal variations in the modular organization of functional networks.
\cite{Betzel2016} focused on periods when time-resolved functional connectivity on a large number 
of edges is stronger or weaker than expected by chance, given its long-timescale functional 
connectivity, and showed that these periods are associated with periods of high modularity in 
time-resolved functional networks.
\cite{Shine2016} defined segregated and integrated states of time-resolved functional networks based 
on fluctuations in nodal participation coefficients and within-module degree \citep{Guimera2005} and 
demonstrated that the modularity of time-resolved networks in the segregated state is greater than 
that in the integrated state.
The latter two studies have established the close relationship of fluctuations in connectivity 
strengths and network topology with fluctuations in modularity of time-resolved functional 
networks.
However, no previous study has directly tracked temporal changes in modularity to characterize 
time-resolved network configurations during periods of high and low modularity.

The present study aims to reveal the following three basic properties of time-resolved functional 
networks in the high and low modularity periods.
(1) {\it Spatial connectivity patterns}.
What connectivity patterns are dominant in the high (low) modularity period?
To clarify this point, we extract periods of high and low modularity in time-resolved functional 
networks from human rs-fMRI data and characterize patterns in time-resolved connectivity 
averaged within the high or the low modularity period.
We also compare these patterns to those obtained using methods previously employed to 
extract a low-dimensional representation of the time-resolved connectivity, the $k$-means clustering 
algorithm \citep{Allen2014} and principal component analysis (PCA; \citealp{Leonardi2013}).
(2) {\it Temporal homogeneity}.
How homogeneous are different instances of time-resolved functional connectivity sampled from within 
the high (low) modularity period?
To answer this question, we examine the similarity of time-resolved functional connectivity among 
within-period samples and their temporal average.
This analysis helps to explain how time-resolved samples of functional connectivity shape the 
long-timescale correlation structure in rs-fMRI data and the averaged connectivity patterns in the 
high or the low modularity period.
(3) {\it Test-retest reliability}.
Are there differences in the rate of occurrence of the high and low modularity periods across 
individuals?
If so, how consistently do the high and low modularity periods appear within each individual?
To address these questions, we compute frequency and mean dwell time of the high (low) modularity 
period in each individual and then assess their individual differences and test-retest reliability 
using multi-session rs-fMRI data.
The test-retest analysis is also performed on the modularity of long-timescale functional 
connectivity so that we can interpret individual variations in modularity reported in previous 
studies from a time-resolved network perspective.

\section{Materials and Methods}
\subsection{Dataset}
The primary dataset in this study comes from the Washington University-University of Minnesota 
(WU-Minn) consortium of the Human Connectome Project (HCP; \url{http://www.humanconnectome.org}).
We also employed another independent dataset from the enhanced Nathan Kline 
Institute-Rockland Sample (NKI-RS; \url{http://fcon_1000.projects.nitrc.org/indi/enhanced}) to 
demonstrate the reproducibility of our findings.
Details about the HCP dataset are described in this section while the NKI dataset is detailed in 
{\bf Supplementary Methods}.

Participants in the HCP dataset were recruited by the WU-Minn HCP consortium and all subjects 
provided written informed consent \citep{VanEssen2013}.
Instead of using larger samples from the more than $1{,}000$ subjects in the HCP, we focused on the 
sample set labeled {\it 100 Unrelated Subjects} in the database of the HCP (ConnectomeDB, 
\url{https://db.humanconnectome.org}), since the detection of communities in time-resolved 
functional networks in this study is computationally demanding.
This sample set has been used in previous studies investigating connectivity states of 
time-resolved functional networks \citep{Shine2016,Shine2016a} and evaluating the 
test-retest reliability of graph-theoretical measures \citep{Termenon2016}.
From these $100$ unrelated subjects, we excluded $15$ participants because of their head movements 
during rs-fMRI scans, which met at least one of the following criteria in any of runs 
\citep{Xu2015}: (1) maximum translation $>3$ mm, (2) maximum rotation $>3^\circ$ or (3) mean 
framewise displacement (FD) $>0.2$ mm, where the FD was computed using the $l$2 norm of the six 
translation and rotation parameter differences in motion correction.
We eliminated in addition one participant aged $\geq 36$ years and finally obtained 
a sample of healthy adults aged $\geq$ $22$ years and $<36$ years, comprising $84$ individuals ($40$ 
males).

Imaging data in the HCP dataset were acquired using a modified 3T Siemens Skyra scanner with a 
32-channel head coil.
Resting-state fMRI data in an eyes open condition were collected in four runs of approximately $14$ 
min ($1{,}200$ time points) each, two runs in one session on day 1 and two runs in another session on 
day 2 (scanning parameters: repetition time (TR) $=720$ ms, echo time (TE) $=33.1$ ms, flip angle 
$=52^\circ$, voxel size $=2$ mm isotropic, field of view (FOV) $=208\times180$ mm$\null^2$ and $72$ 
slices).
In each session, the data were acquired with opposing phase encoding directions, left-to-right (LR) 
in one run and right-to-left (RL) in the other run.
Scanning parameters of a T1-weighted structural image were TR $=2{,}400$ ms, TE $=2.14$ ms, flip angle 
$=8^\circ$, voxel size $=0.7$ mm isotropic, FOV $=224\times224$ mm$\null^2$ and $320$ slices.

We used images in ConnectomeDB preprocessed with the minimal preprocessing pipelines adopted by the 
HCP \citep{Glasser2013}.
The minimal preprocessing for rs-fMRI data included (1) gradient distortion correction, (2) motion 
correction, (3) bias field removal, (4) spatial distortion correction, (5) transformation to 
Montreal Neurological Institute (MNI) space and (6) intensity normalization.
The rs-fMRI data were further preprocessed in the following order: (1) discarding of the first $10$ 
s volumes, (2) removal of outlier volumes and interpolation, (3) nuisance regression using global, 
white matter and cerebrospinal fluid (CSF) mean signals and the Friston-24 motion time series 
\citep{Friston1996}, (4) temporal band-pass filtering (cutoff: low, $1/(66\ \mathrm{TRs})=0.021$ Hz; 
high, $0.1$ Hz; the low-cut frequency was specified as the reciprocal of the width of the time 
window for computing time-resolved functional connectivity) and (5) linear and quadratic detrending.
Our approach to remove outliers is essentially similar to the motion scrubbing \citep{Power2012} and 
censoring \citep{Power2014}; however, instead of removing affected time points, we replaced outliers 
with an interpolated value to keep the number of time points in a sliding time window.
As in \cite{Allen2014}, outlier detection and interpolation were performed using the function 
{\it 3dDespike} in AFNI \citep{Cox2012}.

\subsection{Cortical parcellation}
Connectivity analyses were performed in a region-wise manner within the cortex.
In the HCP dataset, the cortex was parcellated into $114$ distinct regions on the basis of a 
subdivision of the Desikan-Killiany anatomical atlas in FreeSurfer \citep{Cammoun2012} (see 
Supplementary Fig.~\ref{fig:rois_lausanne}).
The subdivision was performed using the atlas files {\it myatlas\_60\_lh.gcs} and 
{\it myatlas\_60\_rh.gcs} in Connectome Mapper (\url{https://github.com/LTS5/cmp}).
Based on the area of overlap, we associated every parcel with one of the $7$ network components, 
the control network (CON), the default mode network (DMN), the dorsal attention network (DAN), the 
limbic system (LIM), the saliency/ventral attention network (VAN), the somatomotor network (SMN) and 
the visual network (VIS), in a functional cortical parcellation defined based on the similarity 
of intrinsic functional connectivity profiles in $1{,}000$ subjects \citep{Yeo2011}.

\subsection{Functional connectivity and window parameters}
As a metric of functional connectivity, we used the Fisher $z$-transformed Pearson correlation 
coefficient between pairs of rs-fMRI time series averaged within each cortical region.
The raw correlation coefficient was used only when presenting a functional connectivity matrix in 
figures.
The conventional long-timescale functional connectivity was derived from the total duration of 
rs-fMRI time series.
In contrast, the time-resolved functional connectivity was computed using a tapered sliding window 
approach.
Window parameters were specified such that the shape of tapered window and the between-window 
duration became similar to those employed in \cite{Allen2014}.
Specifically, the tapered window was created by convolving a rectangle (width $=66$ TRs $=47.52$ s) 
with a Gaussian kernel ($\sigma=9$ TRs $=6.48$ s) and was moved in steps of $3$ TRs $=2.16$ s, 
resulting in a total number of $369$ windows.

\subsection{Modularity of functional networks}
A modularity quality function \citep{Newman2004}, from which communities in a network were detected, 
is a widely used measure of the modularity of a network.
Since functional brain networks may contain negative edge weights, we adopted the asymmetric 
generalization of the modularity quality function introduced in \cite{Rubinov2011}:
\begin{align}\label{eq:modularity}
 Q=\frac{1}{\nu^{+}}\sum_{i,j}\left(w_{i,j}^{+}-e_{i,j}^{+}\right)\delta_{M_i,M_j}
   -\frac{1}{\nu^{+}+\nu^{-}}\sum_{i,j}\left(w_{i,j}^{-}-e_{i,j}^{-}\right)\delta_{M_i,M_j},
\end{align}
where $w_{i,j}^{+}$ is equal to the $i,j$-th entry of a functional connectivity matrix $w_{i,j}$ if 
$w_{i,j}>0$ and is equal to zero otherwise.
Similarly, $w_{i,j}^{-}=-w_{i,j}$ if $w_{i,j}<0$ and $w_{i,j}^{-}=0$ otherwise.
The term $e_{i,j}^{\pm}=s_{i}^{\pm}s_{j}^{\pm}/\nu^{\pm}$ stands for the expected density of 
positive or negative edge weights given a strength-preserved random null model, where 
$s_{i}^{\pm}=\sum_{j}w_{i,j}^{\pm}$ and $\nu^{\pm}=\sum_{i,j}w_{i,j}^{\pm}$.
The Kronecker delta function $\delta_{M_i,M_j}$ is equal to one when the $i,j$-th nodes are in the 
same community and is equal to zero otherwise.

Communities in functional networks were detected by maximizing this quality function $Q$ using the 
Louvain algorithm \citep{Blondel2008} and the maximized $Q$ value was used as a measure of 
modularity evaluating the goodness of the resulting partition.
Community detection and $Q$ maximization were performed using the function {\it community\_louvain.m} 
in the Brain Connectivity Toolbox (BCT version 2016-01-16; 
\url{http://www.brain-connectivity-toolbox.net}) with the default resolution parameter $\gamma=1$.
We ran this function $100$ times and chose the maximum $Q$ and its corresponding partition 
for further analyses.
For time-resolved functional networks, the quality function was computed separately in each time 
window, yielding time-resolved partitions and modularity scores $Q_t$ for $t=1,\ldots,T$, where 
$T$ is the number of windows.
It should be mentioned that both $Q$ and $Q_t$ cannot be properly compared between the HCP and the 
NKI datasets, since cortical parcellations employed in these datasets were different from each other 
while the behavior of these quality functions depend both on the size of networks as well as the 
resolution of module partitions \citep{Good2010}.

\subsection{Periods of high and low modularity}
\begin{figure}[tb]
  \centering
  \includegraphics[scale=0.62]{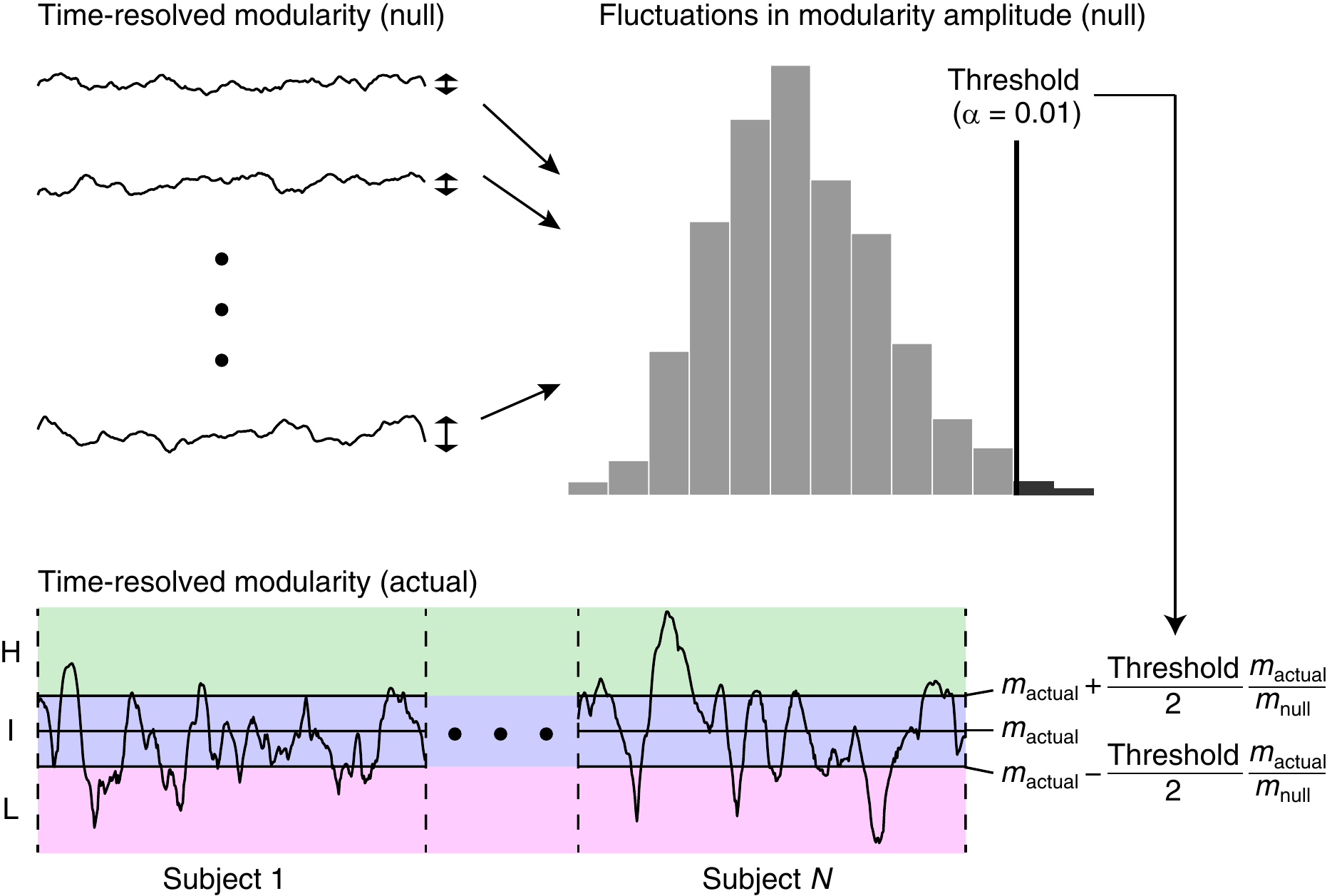}
\caption
 {The procedure to determine the high and low modularity periods.
 Modularity of null time-resolved functional connectivity was generated to obtain null distributions 
 of fluctuations in modularity amplitude.
 The threshold with $\alpha=0.01$ in this null distribution was used for determining periods 
 when actual time-resolved modularity was significantly greater or smaller than its median across 
 subjects $m_{\mathrm{actual}}$.
 With this threshold and the normalization constant $m_{\mathrm{actual}}/m_{\mathrm{null}}$ 
 ($m_{\mathrm{null}}$: the median of time-resolved modularity across all samples from the null 
 distribution), time-resolved functional connectivity in each time window was associated with either 
 the high modularity (H), the intermediate (I) or the low modularity (L) period (also see the 
 paragraph containing Eqs.~(\ref{eq:hiQ}) and (\ref{eq:loQ})).}
\label{fig:workflow}
\end{figure}

Periods of high and low modularity in time-resolved functional connectivity were defined as periods 
in which $Q_t$ was significantly greater and smaller than its median across all subjects, 
respectively (see Fig.~\ref{fig:workflow}).
Here we determined the high and low modularity periods across individuals in order to characterize 
time-resolved network properties during periods of high and low modularity in an absolute sense.
The statistical significance was evaluated based on null distributions of fluctuations in modularity 
amplitude, derived from null models assuming stationarity of time-resolved functional connectivity.

Null time-resolved functional connectivity was generated using stationary vector autoregressive (VAR) 
models \citep{Chang2010,Zalesky2014,DePasquale2016,Shine2016,Shine2016a}.
Parameters in the null stationary VAR models were estimated from actual rs-fMRI time series for 
each pair of regions.
The order of the VAR models was set to $11$ (in the HCP dataset) and $12$ (in the NKI dataset) to 
use a maximum lag of approximately $8$ s \citep{Zalesky2014,Shine2016}.
The estimated VAR models were used for individually simulating null stationary rs-fMRI time series 
for each pair of regions, where initial values for the simulation were a randomly-sampled contiguous 
block of actual rs-fMRI time series and the innovation term was a randomly-sampled residual of the 
VAR estimation.
From the simulated rs-fMRI data the null time-resolved functional connectivity was computed 
in the same way as the actual data.
As in \cite{Zalesky2014} and \cite{Shine2016,Shine2016a}, a total of around $2{,}500$ null samples 
(HCP, $30$ samples $\times 84$ subjects; NKI, $32$ samples $\times 80$ subjects) were generated from 
actual data of all subjects for each rs-fMRI run.

Null distributions of fluctuations in modularity amplitude were obtained from $Q_t$ of null 
time-resolved functional connectivity.
A threshold with $\alpha=0.01$ in this null distribution (see Fig.~\ref{fig:workflow}) was used to 
determine the high and low modularity periods.
The high modularity period was defined as the period when
\begin{align}\label{eq:hiQ}
  Q_t>m_{\mathrm{actual}}+\frac{\mathrm{Threshold}}{2}\frac{m_{\mathrm{actual}}}{m_{\mathrm{null}}}
\end{align}
holds, where $m_{\mathrm{actual}}$ and $m_{\mathrm{null}}$ are the median of actual and null 
time-resolved modularity across all subjects and samples, respectively.
Similarly, the low modularity period was defined as the period when
\begin{align}\label{eq:loQ}
  Q_t<m_{\mathrm{actual}}-\frac{\mathrm{Threshold}}{2}\frac{m_{\mathrm{actual}}}{m_{\mathrm{null}}}
\end{align}
holds and the intermediate period was the period other than the high or the low modularity period.
In Eqs.~(\ref{eq:hiQ}) and (\ref{eq:loQ}) we multiplied the threshold by the normalization term 
$m_{\mathrm{actual}}/m_{\mathrm{null}}$ to absorb potential discrepancies in overall $Q_t$ 
magnitude between the actual data and the null data generated from the pairwise VAR simulations.

We generated null data from stationary VAR models to determine the periods of high and low 
modularity because these null models have been used previously in studies by \cite{Zalesky2014} and 
\cite{Shine2016,Shine2016a} that examined fluctuations in network attributes closely related to 
modularity, i.e.\ efficiency and the balance of integration and segregation.
Our use of VAR models allows a more direct comparison of our own findings with those of these 
earlier studies.
There are other approaches to generate null data, for instance, phase randomization with the same 
random sequence of phases \citep{Prichard1994} and generating covariance- and spectra-constrained 
random normal deviates \citep{Laumann2016}.
Recent discussion on the use of null models in time-resolved functional connectivity studies can be 
found in \cite{Liegeois2017} and \cite{Miller2017}.

 \subsection{Analysis of spatial connectivity patterns}
Spatial patterns of time-resolved functional connectivity in the high or the low modularity period 
were mainly assessed with their temporal median (i.e., the centroid of time-resolved functional 
connectivity in the high or the low modularity period).
These averaged connectivity patterns were characterized by referring to known intrinsic functional 
connectivity networks \citep{Yeo2011} as well as by comparing to the connectivity patterns 
observed in low-dimensional functional connectivity states detected by the $k$-means clustering 
algorithm and PCA\null.

The $k$-means clustering algorithm has been widely used for detecting a small set of brief 
functional connectivity patterns 
\citep{Allen2014,Damaraju2014,Barttfeld2015,Hutchison2015,Gonzalez-Castillo2015,Rashid2016,Nomi2016}.
We performed $k$-means clustering using the same procedure as in \cite{Allen2014} and 
\cite{Barttfeld2015}.
In this procedure time-resolved functional connectivity concatenated across subjects 
was subsampled along the time dimension before clustering of all samples.
The subsampling was performed by choosing local maxima of the time series of functional connectivity 
variance, resulting in $15$--$35$ (HCP) and $24$--$41$ (NKI) windows per subjects in subject 
exemplars.
The clustering algorithm was first applied to this set of subject exemplars $500$ times with the 
$l$1 distance metric and random initialization, and then the obtained median centroids with the 
minimum error were used as an initial starting point for the subsequent $k$-means clustering of all 
time window samples.
We used $k=2$ as the number of clusters in order to extract two major distinctive connectivity 
states, and we compared the connectivity patterns observed in their cluster centroids to the 
patterns found in the centroids of the high and low modularity periods.

Unlike the $k$-means clustering, PCA has been used to extract a low-dimensional representation of 
time-resolved functional connectivity in a continuous manner.
We performed PCA according to the approach described in \cite{Leonardi2013}.
The principal components were obtained by eigenvalue decomposition: $\vm C\vm C^{\te}=\vm U\vm D\vm 
U^{\te}$, where the matrix $\vm C$ denotes time-resolved functional connectivity concatenated across 
subjects (here the temporal average of time-resolved connectivity was subtracted individually).
The matrix $\vm U$ contains the eigenvectors of $\vm C\vm C^{\te}$ (i.e., the principal components 
of $\vm C$) in its columns and $\vm D$ is a diagonal matrix containing the corresponding eigenvalues.
The weights of principal components were derived as $\vm U^{\te}\vm C$, representing the contribution 
of each principal component in the variability of time-resolved functional connectivity over time.
We mainly focused on the principal component with the largest eigenvalue (the first principal 
component) and compared the connectivity patterns in the first principal component to the patterns 
observed in differences of centroid edge weights between the high and low modularity periods.

\subsection{Analysis of temporal homogeneity}
Temporal homogeneity of time-resolved functional connectivity within the high or the low modularity 
period was evaluated using similarity measures for edge weights and partitions of functional networks.
These similarity measures were computed between samples of time-resolved functional connectivity 
in each modularity period within individuals, as well as between each sample of time-resolved 
functional connectivity and its corresponding centroid.
The similarity of edge weights was quantified by the Pearson correlation coefficient between the 
vectorized weights.
The similarity of partitions was quantified by the normalized mutual information using the BCT 
function {\it partition\_distance.m}.

\subsection{Analysis of test-retest reliability}
Test-retest reliability of the occurrence of the high and low modularity periods within each 
individual was evaluated using the multi-session rs-fMRI data in the HCP dataset.
The occurrence of these periods was assessed through their frequency and mean dwell time 
\citep{Allen2014,Damaraju2014,Hutchison2015}.
The frequency is defined as the ratio of the number of time windows classified into a particular 
modularity period relative to the total number of windows.
The mean dwell time is the number of consecutive windows classified into a particular modularity 
period, averaged within each subject.
The mean dwell time of a modularity period to which no windows were assigned was set to zero.
We omitted the first and last segments of consecutive windows from the mean dwell time since the 
beginnings and/or the ends of these episodes fall outside of the scanning period.
We did not define the mean dwell time of a modularity period in a subject when all its consecutive 
windows overlapped with either the beginning or the end of the entire scan period.

The test-retest reliability was quantified by the intra-class correlation coefficient (ICC; 
\citealp{Shrout1979}).
The ICC has been used in a number of test-retest studies on rs-fMRI data 
(\citealp{Schwarz2011,Wang2011,Braun2012,Guo2012,Cao2014a,Andellini2015,Termenon2016}; for a review, 
see \citealp{Zuo2014}).
We computed the ICC under a two-way mixed model \citep{McGraw1996}:
\begin{align}\label{eq:icc}
 \mathrm{ICC}=\frac{\mathrm{BMS}-\mathrm{EMS}}{\mathrm{BMS}+(l-1)\mathrm{EMS}},
\end{align}
where $\mathrm{BMS}$ is the between-subjects mean square, $\mathrm{EMS}$ is the error mean square 
and $l$ is the number of repeated observations per subject.
According to previous rs-fMRI test-retest studies \citep{Guo2012,Andellini2015}, we interpreted 
$0.2<\mathrm{ICC}\leq 0.4$ as indicative of a fair test-retest reliability and $\mathrm{ICC}>0.4$ 
as moderate to good test-retest reliability.
Consistent with \cite{Braun2012}, \cite{Cao2014a} and \cite{Andellini2015}, negative ICC scores were 
set to zero, since the reason for the presence of negative ICCs is unclear \citep{Muller1994} and 
negative reliability is difficult to interpret \citep{Rousson2002}.

\section{Results}
Most of the results presented in this section (in Figs.~\ref{fig:qstates}--\ref{fig:pca} and Table 
\ref{tab:correlation}) are derived from run 2LR of the HCP dataset.
Reproducibility of the findings about the spatial connectivity patterns and the temporal homogeneity 
were examined in {\bf Supplementary Results} using the other three runs of the HCP dataset (and a 
single run of the NKI dataset).
We selected run 2LR because time-resolved modularity $Q_t$ in this run was least affected by head 
movements; the correlation coefficient between $Q_t$ and the sliding-window-averaged FD, 
averaged over subjects, was $-0.11$ in run 1LR, $-0.056$ in run 1RL, $-0.029$ in run 2LR and $-0.10$ 
in run 2RL.
This mean correlation coefficient was also found to be small in the NKI dataset ($-0.064$).
The relationships of $Q_t$ to FD are further evaluated in more detail in {\bf Supplementary Results}.

\subsection{Spatial connectivity patterns}
\begin{figure}[tb]
  \centering
  \includegraphics[scale=0.62]{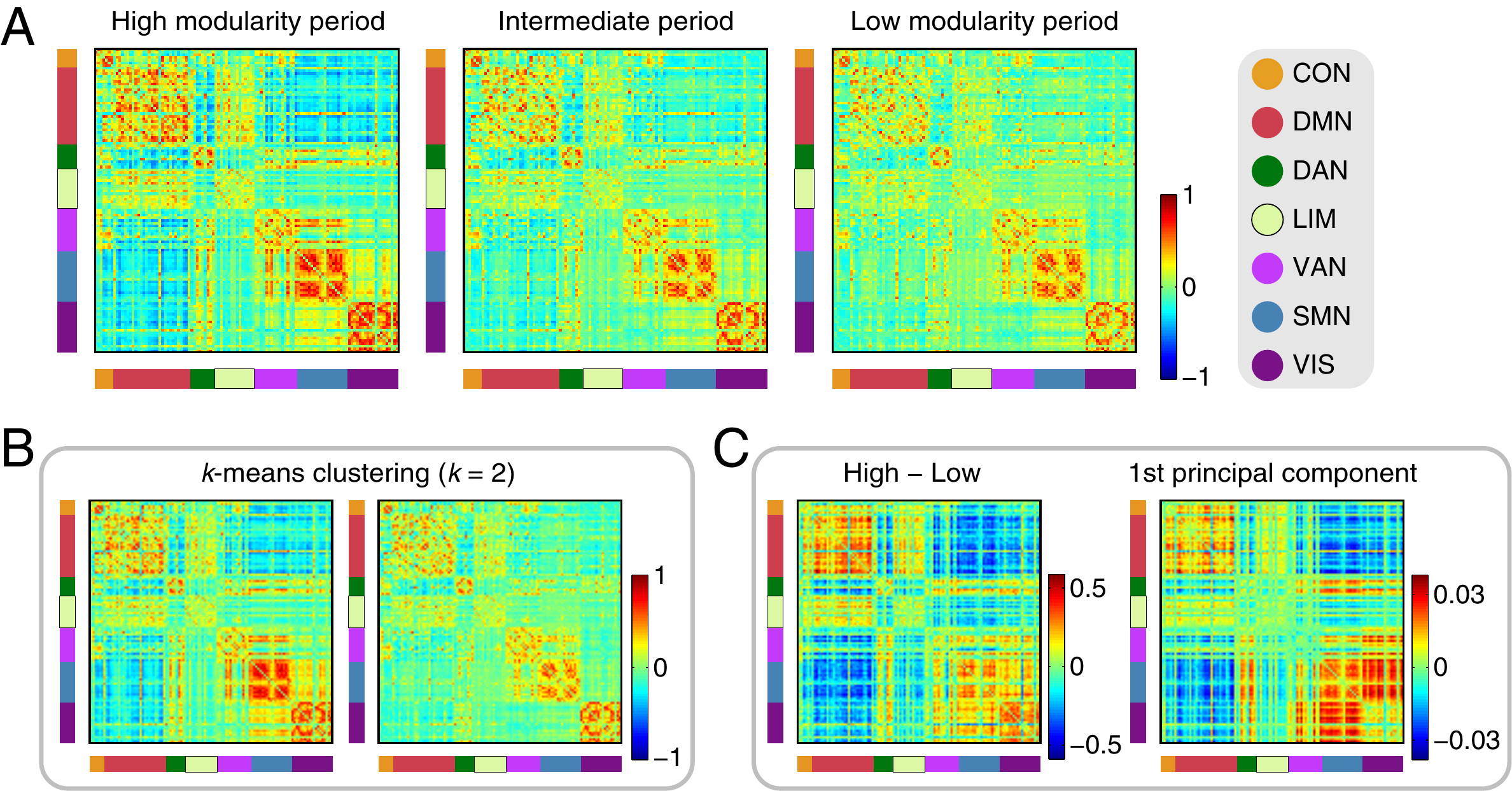}
\caption
 {{$\vm A$}, Centroids of time-resolved functional connectivity in the high modularity, the 
 intermediate and the low modularity periods.
 A list of the assignment of cortical nodes to the $7$ network components in \cite{Yeo2011} is 
 presented in Supplementary Fig.~\ref{fig:rois_lausanne}.
 {$\vm B$}, Centroids derived from the $k$-means clustering algorithm with $k=2$.
 {$\vm C$}, Left: differences of centroid edge weights between the high and low modularity periods.
 Right: the first principal component of time-resolved functional connectivity.}
\label{fig:qstates}
\end{figure}

First, we examined spatial patterns of time-resolved functional connectivity averaged within each 
modularity period.
Figure \ref{fig:qstates}A shows the centroids of the high modularity, the intermediate and the low 
modularity periods.
The centroid of the high modularity period had the highest contrast of edge weights, exhibiting the 
modular organization of distinct functional networks most clearly.
Differences of centroid edge weights between the high and low modularity periods can be 
characterized as a pronounced dissociation of the task-negative DMN module from the task-positive 
DAN, VAN, SMN and VIS modules in the high modularity period.
These spatial patterns were less evident in the low modularity period, which exhibited much less 
differentiated (''flat'') connectivity patterns in its centroid.
Increased dissociation of the task-positive and task-negative systems in the high modularity period 
can also be seen in Supplementary Fig.~\ref{fig:reordered_centroids}, where nodes in the centroids 
are reordered based on communities detected by modularity maximization in long time-scale functional 
connectivity.

We then compared the connectivity patterns observed in the centroids of the high and low modularity 
periods to the patterns derived from the $k$-means clustering algorithm and PCA\null.
Figure \ref{fig:qstates}B shows that the centroids of the high and low modularity periods had similar 
connectivity patterns to the patterns observed in the centroids of the two mutually-distinctive 
connectivity states detected using $k$-means clustering with $k=2$.
Differences of centroid edge weights between the high and low modularity periods also resembled the 
first principal component of time-resolved functional connectivity (see Fig.~\ref{fig:qstates}C), 
indicating that transitions between the high and low modularity periods mainly occurred along the 
direction of the principal component that explained the largest amount of variance.
These results suggest that transitions between the high and low modularity periods, accompanied by 
changes in the degree of dissociation of the DMN from task-positive networks, accounted for much of 
the observed fluctuations in time-resolved functional connectivity.

\subsection{Temporal homogeneity}
Next, we investigated temporal homogeneity of time-resolved functional connectivity within each 
modularity period.
We first looked into the similarity of time-resolved connectivity in a single representative subject.
Then, we assessed the similarity of time-resolved connectivity using the data from all subjects.

\begin{figure}[b]
  \centering
  \includegraphics[scale=0.62]{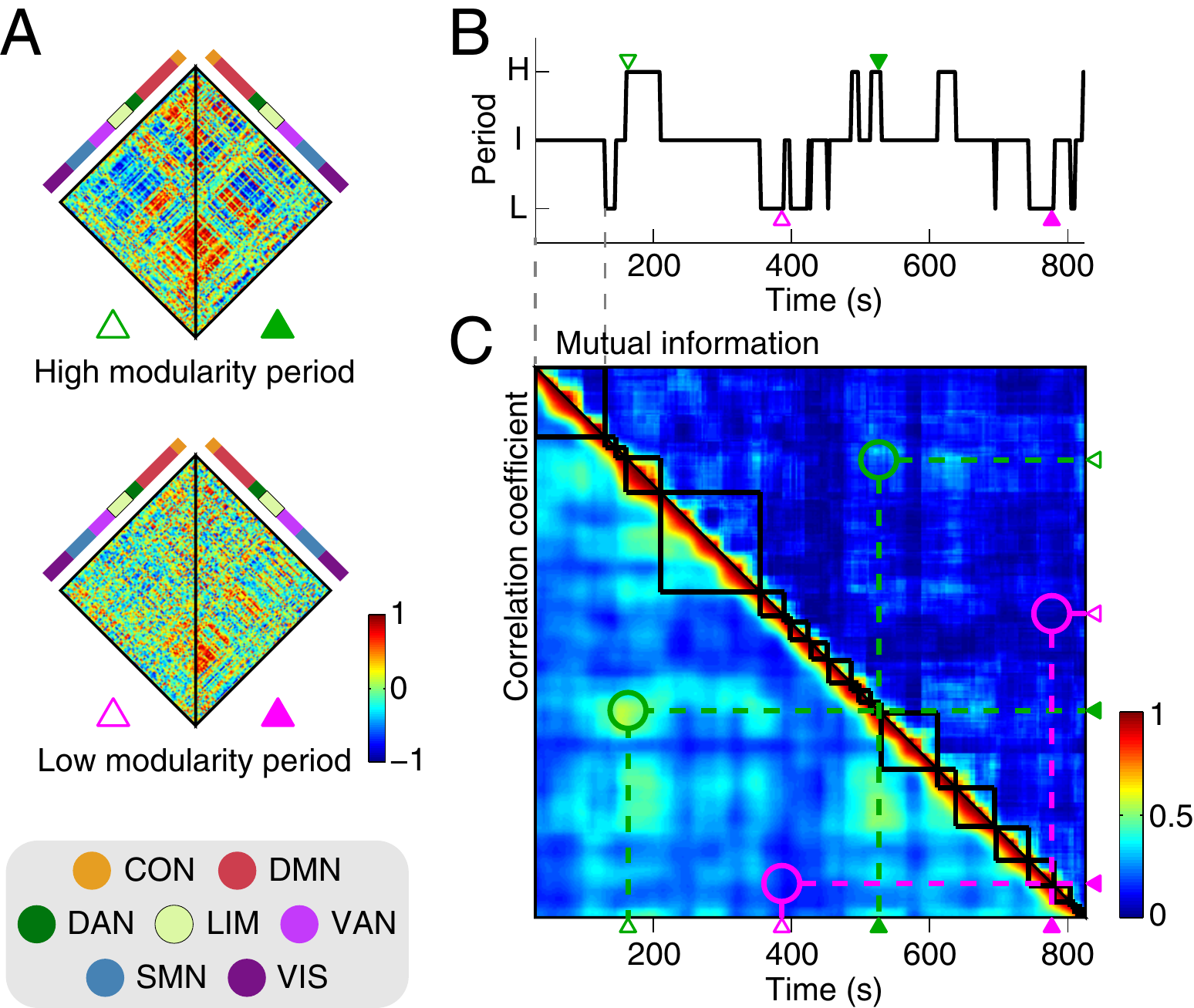}
\caption
 {Analysis of temporal homogeneity in one representative subject (subject ID in the HCP database: 
 133928).
 $\vm A$, Exemplars of time-resolved functional connectivity in the high and low modularity periods.
 $\vm B$, Transition sequence of modularity periods.
 The green and magenta triangles are placed at the positions of the time windows from which 
 the exemplars shown in $\vm A$ were computed.
 $\vm C$, Similarity between samples of time-resolved functional connectivity.
 The lower triangular part shows the similarity of edge weights quantified by the Pearson 
 correlation coefficient and the upper triangular part presents the similarity of partitions 
 quantified by the normalized mutual information.
 The green and magenta circles highlight the similarity scores between the exemplars presented in 
 $\vm A$.}
\label{fig:similarity_eg}
\end{figure}

\subsubsection{Individual subject analysis}
Figure \ref{fig:similarity_eg}A displays exemplars of time-resolved functional connectivity in the 
high and low modularity periods.
As in the centroids shown in Fig.~\ref{fig:qstates}A, the dissociation of the DMN module from the 
task-positive network modules was more evident in exemplars sampled from the high modularity period, 
whereas this module configuration was less pronounced in exemplars sampled from the low 
modularity period.
These exemplars of time-resolved functional connectivity were computed from the time windows whose 
temporal positions are indicated by the small triangles in the transition sequence of modularity 
periods shown in Fig.~\ref{fig:similarity_eg}B\null.
Figure \ref{fig:similarity_eg}C shows the matrix displaying the similarity between samples of 
time-resolved functional connectivity, where the same time axis is used in 
Fig.~\ref{fig:similarity_eg}B and C\null.
This similarity matrix demonstrates that samples of time-resolved functional connectivity in the 
high modularity period were similar to each other, whereas those in the low modularity period were 
dissimilar to each other (the similarity of the exemplars presented in Fig.~\ref{fig:similarity_eg}A 
is highlighted by the circles in Fig.~\ref{fig:similarity_eg}C).

\subsubsection{Group analysis}
\begin{figure}[b]
  \centering
  \includegraphics[scale=0.62,angle=90]{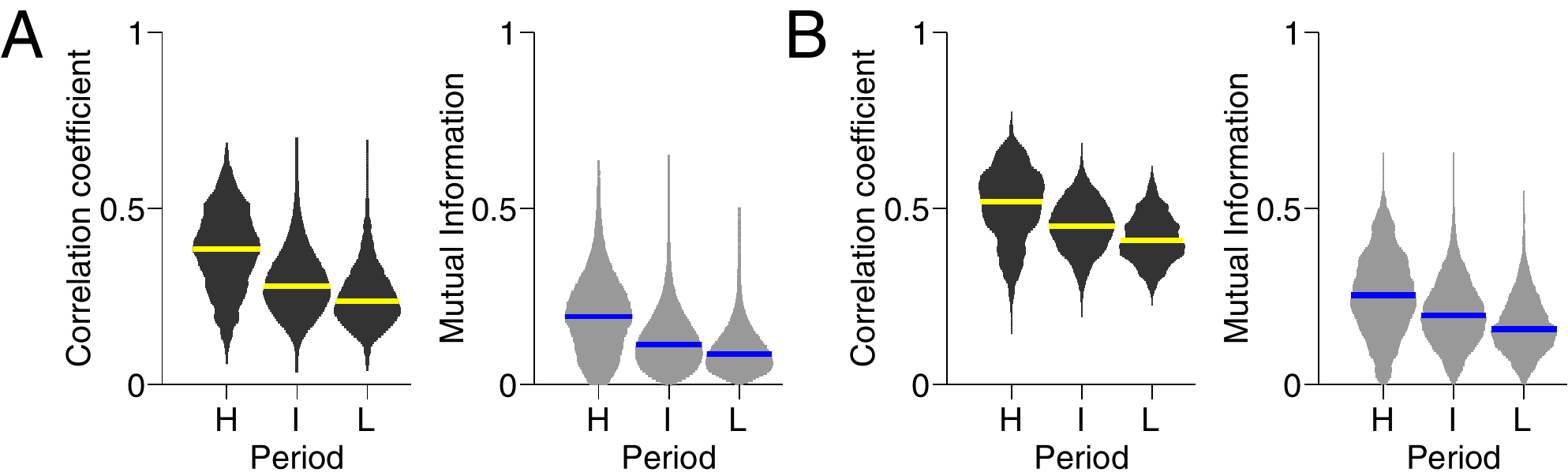}
\caption
 {Distributions of the similarity measures of time-resolved functional connectivity.
 A bar in a distribution indicates the median.
 {$\vm A$}, The similarity of edge weights (left; Pearson correlation coefficient) and partitions 
 (right; normalized mutual information) between samples of time-resolved functional connectivity 
 in each modularity period within individuals.
 These distributions were derived only from pairs of samples that are apart from each other by 
 more than the width of the time window.
 {$\vm B$}, The similarity of edge weights (left) and partitions (right) between each sample of 
 time-resolved functional connectivity and its corresponding centroid.}
\label{fig:violinplots}
\end{figure}

Figure~\ref{fig:violinplots} shows distributions of the similarity scores over all subjects.
Consistent with our observations in the individual subject analysis, samples of time-resolved 
functional connectivity in the high modularity period were more similar to each other compared to 
the low modularity periods, in terms of both edge weights (Cohen's $d=1.4$) and partitions ($d=1.2$) 
(see Fig.~\ref{fig:violinplots}A; the $p$-values were essentially zero due to a large number of 
samples in the distributions).
The greater similarity in the high modularity period was also confirmed when the similarity scores 
were computed between each sample of time-resolved functional connectivity and its corresponding 
centroid, both for edge weights ($d=1.1$) and partitions ($d=0.9$) (see Fig.~\ref{fig:violinplots}B).
These results indicate that the modular connectivity patterns observed in the high modularity 
period are relatively homogeneous in time, while the flat connectivity patterns in the low modularity 
period are relatively heterogeneous.

\begin{figure}[tb]
  \centering
  \includegraphics[scale=0.62]{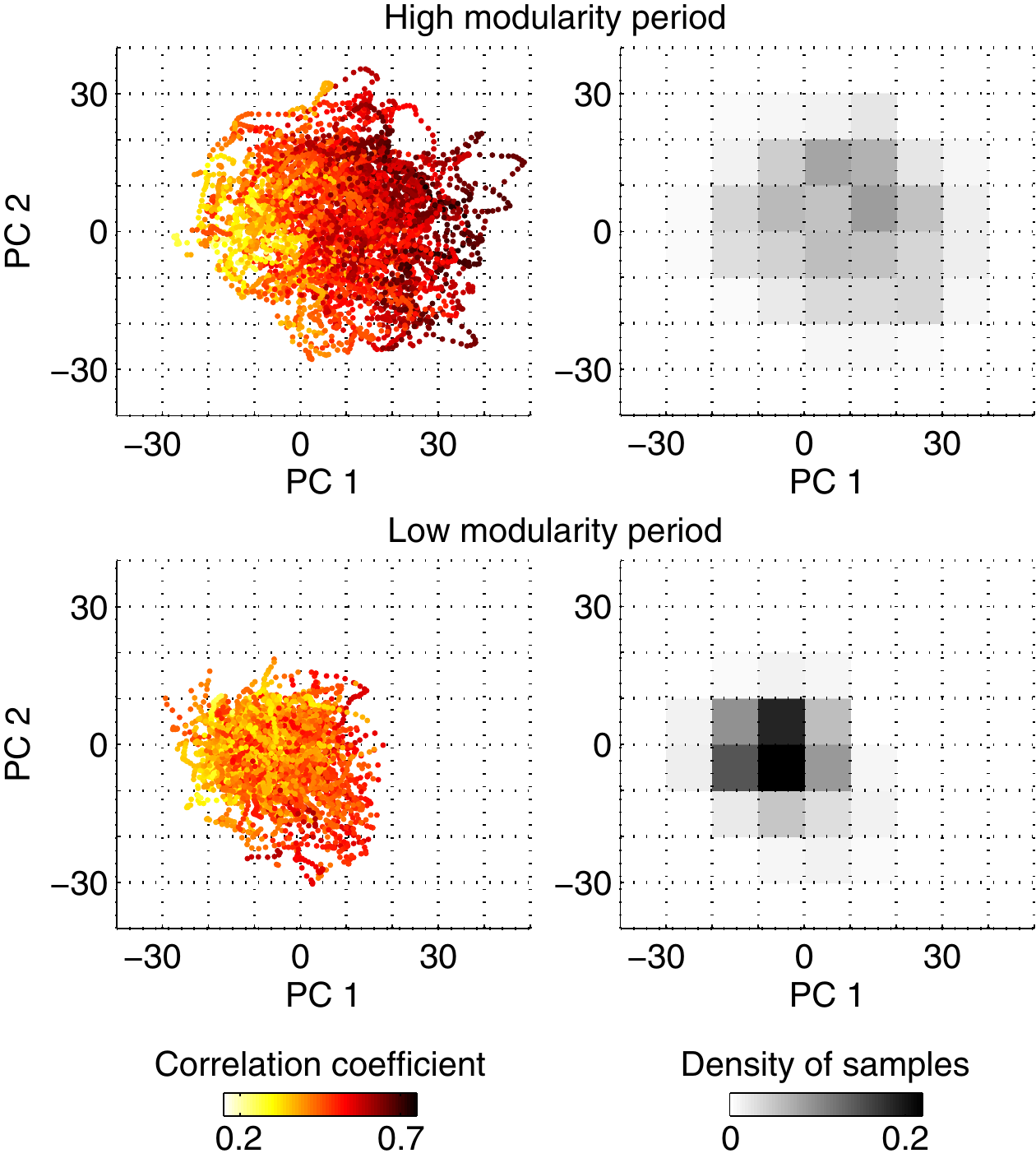}
\caption
 {Left panels: samples of time-resolved functional connectivity in the space spanned by the first 
 and second principal components (PCs 1 and 2).
 Each colored dot represents a sample of time-resolved functional connectivity and its color shows 
 the similarity (Pearson correlation coefficient) of edge weights to its corresponding centroid.
 The numbers of samples in the high and low modularity periods were almost the same ($5{,}329$ and 
 $5{,}310$, respectively).
 Right panels: two-dimensional histograms showing the density of samples of time-resolved functional 
 connectivity in the PC space.}
\label{fig:pca}
\end{figure}

The correlation coefficients displayed in Fig.~\ref{fig:violinplots}B are projected 
into the space spanned by the first and second principal components (PCs 1 and 2) in 
Fig.~\ref{fig:pca} (left panels).
In this figure each colored dot represents a sample of time-resolved functional connectivity and its 
color indicates the similarity to its corresponding centroid.
The color gradient in this figure shows that the greater similarity to respective centroids 
was associated with a larger weight of PC 1, which corresponds to a greater expression of the 
modular connectivity patterns in Fig.~\ref{fig:qstates}C (right) in each sample.
Although the cluster of samples in the high modularity period partially overlapped with the cluster 
of the low modularity period, their centers of mass were separated along the direction of PC 1 
(see Fig.~\ref{fig:pca}, right panels), as is expected from the similarity of connectivity 
patterns between PC 1 and the differences of centroid edge weights (see Fig.~\ref{fig:qstates}C).

It should be mentioned that the homogeneity of time-resolved functional connectivity is not 
necessarily linked to the cohesion of samples in the PC space.
On the contrary, relatively homogeneous samples in the high modularity period were not well clustered 
but rather dispersed, while relatively heterogeneous samples in the low modularity period were 
more densely clustered near the origin of the PC space (see Fig.~\ref{fig:pca}).
This is because, in the low modularity period, a variety of combinations of components other than 
PC 1 or PC 2 shaped heterogeneous connectivity patterns across the clustered samples near the origin.
On the other hand, dispersions to the positive direction of PC 1 resulted in an increased 
homogeneity of samples in the high modularity period, since its centroid and PC 1 shared similar 
connectivity patterns as presented in Fig.~\ref{fig:qstates}.

\subsection{Test-retest reliability}
Finally, we evaluated the test-retest reliability of the occurrence of each modularity period within 
individuals.
We first confirmed that the occurrence of the modularity periods varied across individuals in run 
2LR; the ranges of the frequency across subjects were  $0$--$0.49$ (H), $0.28$--$0.91$ (I) and 
$0.0054$--$0.72$ (L); and the ranges of the mean dwell time were $0$--$62$ s (H), $26$--$114$ s (I) 
and $3.6$--$72$ s (L).
The following test-retest analysis examined how consistently each modularity period appeared within 
individuals over the two sessions ($2\times 28$ min) and the four runs ($4\times 14$ min) of rs-fMRI 
scans in the HCP dataset.

Table \ref{tab:test-retest} displays the ICC of the frequency and mean dwell time of each modularity 
period.
Moderate session-level reliability (ICC $>0.4$) was observed in the frequency of the high 
and low modularity periods as well as the mean dwell time of the low modularity period.
These metrics of occurrence also demonstrated fair (ICC $>0.2$) to moderate run-level 
reliability, while the dwell time of the high modularity period demonstrated only fair session-level 
reliability.
The frequency and mean dwell time of the intermediate period were not reproducible both at the 
session and run levels.
Our test-retest analysis showed that the individual differences in the occurrence of the high and 
low modularity periods were somewhat reproducible.

\begin{table}[b]
 \centering
\caption
 {The ICCs of the frequency and mean dwell time of each modularity period and the ICC of the 
 modularity $Q$ of long-timescale functional connectivity.}\vspace{0mm}
\label{tab:test-retest}
\footnotesize
 \begin{tabular}{cccccccc}
  \toprule
   & \multicolumn{2}{c}{High modularity period} & \multicolumn{2}{c}{Intermediate period} & \multicolumn{2}{c}{Low modularity period} & Modularity $Q$\\
  \cmidrule{2-7}
   & Frequency & Dwell time & Frequency & Dwell time & Frequency & Dwell time &\\
  \midrule
 \multicolumn{8}{l}{Across two sessions (one session: $28$ min)}\\
   ICC & $\vm{0.44}$ & $\textit{\vm{0.21}}$ & $0.19$ & $0$ & $\vm{0.52}$ & $\vm{0.55}$ & $\vm{0.46}$\\
   95\% CI & $[0.25,0.59]$ & $[0.00,0.41]$ & $[0,0.38]$ & $[0,0.10]$ & $[0.34,0.66]$ & $[0.39,0.69]$ & $[0.27,0.61]$\\
 \multicolumn{8}{l}{Across four runs (one run: $14$ min)}\\
   ICC & $\textit{\vm{0.28}}$ & $0.11$ & $0.16$ & $0$ & $\vm{0.42}$ & $\textit{\vm{0.31}}$ & $\vm{0.45}$\\
   95\% CI & $[0.17,0.40]$ & $[0.01,0.22]$ & $[0.06,0.28]$ & $[0,0.08]$ & $[0.30,0.53]$ & $[0.20,0.44]$ & $[0.33,0.56]$\\
  \bottomrule
 \end{tabular}
\vskip -\lastskip \vskip 5pt
\caption*{ICC $>0.4$ is shown in bold and $0.2<$ ICC $\leq 0.4$ is shown in bold italic.
 As explained in the section entitled {\it Analysis of test-retest reliability}, negative ICC scores 
 are regarded as zero.}
\end{table}

The individual differences in the occurrence of the high and low modularity periods were closely 
linked to the previously-reported individual differences in the modularity $Q$ of long-timescale 
functional connectivity.
Consistent with the literature, we observed that the long-timescale modularity $Q$ varied across 
individuals (the range of $Q$ across subjects in run 2LR, $0.37$--$0.64$) with moderate 
session-level and run-level reproducibility (see Table \ref{tab:test-retest}).
Table \ref{tab:correlation} shows that $Q$ was positively (respectively, negatively) correlated to 
the frequency and mean dwell time of the high (respectively, low) modularity period across 
individuals, indicating that the high (low) modular period is more likely to occur and lasts with a 
longer duration in individuals with high (low) $Q$ scores.
This result demonstrated that the individual variations in the modularity $Q$ of long-timescale 
functional connectivity can be explained by the differences in the occurrence of the high and low 
modularity periods across individuals.

\begin{table}[tb]
 \centering
\caption
 {Pearson correlation coefficient between the modularity $Q$ of long-timescale functional connectivity 
 and the frequency and mean dwell time of each modularity period (bold, $p<0.01$).}\vspace{0mm}
\label{tab:correlation}
\footnotesize
 \begin{tabular}{ccccccc}
  \toprule
   & \multicolumn{2}{c}{High modularity state} & \multicolumn{2}{c}{Intermediate state} & \multicolumn{2}{c}{Low modularity state}\\
  \cmidrule{2-7}
   & Frequency & Dwell time & Frequency & Dwell time & Frequency & Dwell time\\
  \midrule
   $r$ & $\vm{0.65}$ & $\vm{0.35}$ & $0.13$ & $-0.03$ & $\bm{-0.65}$ & $\bm{-0.49}$\\
   95\% CI & $[0.50,0.76]$ & $[0.14,0.52]$ & $[-0.08,0.34]$ & $[-0.24,0.19]$ & $[-0.76,-0.50]$ & $[-0.64,-0.31]$\\
  \bottomrule
 \end{tabular}
\end{table}

\section{Discussion}
Although recent studies have reported that fluctuations in edge weights and topology of 
resting-state functional brain networks co-occur with temporal changes in their modularity, the 
spatiotemporal configurations of time-resolved functional networks during periods of high and low 
modularity have remained largely unexplored.
In this study we examined how time-resolved functional networks sampled from periods of high and low 
modularity are organized in space and time, whether the occurrence of these periods varies across 
individuals, and whether individual differences are reproducible across sessions.
Our analysis of spatial connectivity patterns showed that the time-resolved connectivity in the high 
modularity period was characterized by the pronounced dissociation of the DMN from task-positive 
networks and that the transitions between this module structure and the flat connectivity profiles 
of the low modularity period well summarized overall fluctuations of the time-resolved connectivity.
The temporal homogeneity analysis demonstrated that the decoupling of the DMN from task-positive 
networks consistently occurred across samples of the time-resolved connectivity in the high 
modularity period, whereas the flat connectivity patterns observed in the low modularity period were 
relatively heterogeneous and more dissimilar to each other over time.
The test-retest analysis confirmed that the occurrence of the high and low modularity periods varied 
across individuals with moderate inter-session reproducibility and that these variations across 
individuals can explain individual differences in the modularity of long-timescale functional 
connectivity.

Our spatial and temporal analyses of time-resolved functional networks in the high and low 
modularity periods bring a new perspective to previously observed attributes of the correlation 
structure in rs-fMRI data.
It is well-established that time courses of regional activity in the DMN and task-positive 
networks are correlated within and anti-correlated between networks when these correlations 
are computed from several minutes of rs-fMRI measurements \citep{Fox2005,Fox2007}.
Our findings on a time scale of tens of seconds demonstrated that this specific correlation 
structure among the DMN and task-positive networks was recurrently strengthened in the high 
modularity period, whereas this structure was weakened and was broken down heterogeneously in the 
low modularity period.
Temporal fluctuations in expression of this specific correlation structure have already been 
reported in recent studies (e.g., \citealp{Allen2014}; \citealp{Betzel2016}).
In the present paper we demonstrate a link of this structure to variations in the temporal 
homogeneity (or heterogeneity) of short-timescale correlations.
Our results suggest that the canonical long-timescale correlation structure in rs-fMRI data is 
shaped by the repeated appearance of relatively uniform correlation patterns in the high modularity 
period and an averaging out of many dissimilar correlation patterns in the low modularity period.
From this perspective, one can argue that the proportion of the high to the low modularity period 
in rs-fMRI data may underlie the degree to which the DMN is decoupled from task-positive 
networks in estimates of long-timescale correlations.

The findings about the spatial and temporal connectivity profiles in the high and low modularity 
periods also provide another perspective on segregated and integrated states of time-resolved 
functional networks, recently studied by \cite{Shine2016}.
There is growing interest in temporal fluctuations in the balance between segregation and 
integration of neural information \citep{Sporns2013,Deco2015}.
Shine and colleagues defined states of segregation and integration based on patterns in nodal 
participation coefficients and within-module degree of time-resolved functional 
networks and demonstrated that topology of the time-resolved networks in the segregated and 
integrated states was associated with high and low modularity, respectively \citep{Shine2016}.
Importantly, alternations between these segregated and integrated states were related to cognitive 
function and the activity of neuromodulatory systems---a network attribute for integration was found 
to be correlated with measures quantifying fast and effective cognitive performance and with 
increases in pupil diameter as a surrogate measure of arousal \citep{Shine2016}.
We found that the dissociation of the DMN from task-positive networks was diminished and the 
similarity between samples of time-resolved functional connectivity was decreased in the low 
modularity period---hallmarks of network organization that may indeed be related to the level of 
functional integration.
These results suggest that efficient cognitive processing in the integrated state is likely to be 
supported by multiple, heterogeneous coupling patterns across the whole brain, accompanied by 
dynamic reconfiguration of interactions between the DMN and other networks, as seen with greater 
task demands of working memory \citep{Vatansever2015}.

The test-retest analysis of the high and low modularity periods offers a new interpretation of 
individual differences in modularity measured over longer timescales.
Previous studies have shown that this long-timescale modularity varies across individuals and its 
variability is associated with life-span development and aging \citep{Cao2014b,Chan2014}, and 
working memory performance \citep{Stevens2012,Meunier2014,Stanley2014}.
In this study we demonstrated that individual variations in long-timescale modularity are related to 
variations in the occurrence of high- and low-modularity periods, after confirming their moderate 
session-level test-retest reliability.
The greater occurrence of the high (low) modularity period in individuals with higher (lower) 
long-timescale modularity itself is not surprising given their definition; however, it provides a 
way to trace individual variations in long-timescale modularity to the variable occurrence of 
temporal building blocks at shorter time scales.
The relationship established in our test-retest analysis indicates that individual variations in 
long-timescale modularity can be interpreted as a result of variations in the emergence of network 
properties specific to either the high or the low modularity period in time-resolved functional 
connectivity.
The demographic and behavioral relevance of long-timescale modularity may therefore be rooted in the 
differences in the propensity of the DMN to associate with task-positive networks, and in the 
balance between homogeneity and heterogeneity of time-resolved functional connectivity.

There are several methodological limitations in this study.
First, the sliding window approach is a simple and the most commonly used method for computing 
time-resolved functional connectivity \citep{Hutchison2013,Preti2016}, but it has limitations 
in detecting sharp connectivity transitions \citep{Lindquist2014,Shine2015,Shakil2016}.
Since it is necessary to set the width of the time window to be longer than or equal to the 
reciprocal of the lowest frequency of rs-fMRI data to suppress spurious connectivity 
fluctuations \citep{Leonardi2015,Zalesky2015}, changes faster than this specific temporal scale 
cannot be captured with this approach.
This highlights the need to establish methods that can handle connectivity fluctuations on 
multiple time scales.
Second, community detection by modularity maximization fails to detect communities below a certain 
scale under certain conditions (the resolution limit; \citealp{Fortunato2007}).
A single default resolution parameter is used throughout this paper to properly compare the 
modularity quality function; however, this prevents the algorithm from detecting communities of 
different sizes.
A possible strategy to circumvent this issue is to explore a range of resolution parameters to 
reveal multiscale community structure in networks \citep{Betzel2013}.
Third, head movements are a potential confound in rs-fMRI data \citep{Power2012,Power2014}.
To solve this issue, we employed extensive artifact reduction methods in this study; high motion 
subjects were discarded; artifactual BOLD time points were censored and interpolated; motion 
estimates and the global, white matter and CSF mean signals were regressed out.
It should be noted that the findings presented in {\bf Results} (other than in the section {\it 
Test-retest reliability}) were derived from the rs-fMRI run least affected by head movements, and 
that there was no consistent relation between the head movements and the modularity time series in 
this run (see {\bf Supplementary Results}).
And fourth, images in the HCP and the NKI datasets were independently preprocessed.
Nevertheless, the findings in this study (with the exception of the test-retest analysis; the NKI 
dataset contained only a single rs-fMRI run) were reproducible across the HCP and the NKI datasets 
(see {\bf Supplementary Results}).
While there were some minor differences in the spatial connectivity patterns and the temporal 
homogeneity (see {\bf Supplementary Results}), they did not alter the conclusions of this study.

Future research is needed to draw firm conclusions about specific neurobiological mechanisms that 
underlie fluctuations in the modularity of time-resolved functional networks.
For a better understanding of these mechanisms, it is particularly important to establish the 
relationship of dynamic fluctuations in modularity to their underlying structural connectivity.
A simple approach to relate to the structure is evaluating the similarity between structural and 
time-resolved functional connectivity \citep{Barttfeld2015,Liegeois2015}.
This approach can be used to determine in which modularity period time-resolved functional networks 
are more shaped by structural connectivity.
Another promising approach is to simulate fluctuations in modular organization of resting-state 
functional networks using whole-brain computational models constrained by structural connectivity 
\citep{Hansen2015,Ponce2015}.
Generative processes of the modularity fluctuations can be studied by manipulating elements in 
these models and examining its effects on the simulated time-resolved module configuration.
Examining individual differences is also an important direction to relate the modularity 
fluctuations to the underlying structure.
Future analyses integrating individual variations in the occurrence of the high and low modularity 
periods and structural connectivity, in addition to demographics and cognitive performance, may 
contribute to a more comprehensive understanding of fluctuations in the modularity of time-resolved 
functional networks.

\section*{Acknowledgments}
M.F. was supported by a Uehara Memorial Foundation Postdoctoral Fellowship and a Japan Society for 
the Promotion of Science Postdoctoral Fellowship for Research Abroad.
O.S. was supported by the J.S. McDonnell Foundation (22002082) and the National Institutes of Health 
(R01 AT009036-01).
R.F.B. was supported by the National Science Foundation/Integrative Graduate Education and Research 
Traineeship Training Program in the Dynamics of Brain-Body-Environment Systems at Indiana University 
(0903495).
X.N.Z was supported by the National Key Basic Research and Development Program (973 Program; 
2015CB351702) and the Natural Sciences Foundation of China (81471740, 81220108014).
X.N.Z. and O.S. are members of an international collaboration team (trial stage) supported by the 
CAS K.C. Wong Education Foundation.
Data were provided in part by the Human Connectome Project, WU-Minn Consortium (Principal 
Investigators: David Van Essen and Kamil Ugurbil; 1U54MH091657) funded by the 16 NIH Institutes and 
Centers that support the NIH Blueprint for Neuroscience Research; and by the McDonnell Center for 
Systems Neuroscience at Washington University.

\newpage
\thispagestyle{empty}
\section*{\Large{Highlights}}
\begin{itemize}
\item{Time-resolved functional connectivity in high/low modularity periods is investigated.}
\item{High modularity is tied to increased dissociation of task-positive/negative networks.}
\item{Low modularity is associated with heterogeneous connectivity patterns over time.}
\item{Frequency of high/low modularity periods has moderate inter-session reproducibility.}
\item{An interpretation of individual differences in long-timescale modularity is provided.}
\end{itemize}

\newpage
\section*{\Large{Supplementary Material}}
\pagenumbering{roman}
\setcounter{figure}{0}
\setcounter{table}{0}
\renewcommand{\figurename}{Supplementary Figure}
\renewcommand{\tablename}{Supplementary Table}
\makeatletter
  \renewcommand{\thefigure}{%
  S\arabic{figure}}
 \@addtoreset{figure}{section}
  \renewcommand{\thetable}{%
  S\arabic{table}}
 \@addtoreset{table}{section}
\makeatother

\section*{Supplementary Methods}
\subsection*{Replication dataset}
As an independent dataset for replication analyses, we used the NKI dataset employed in our previous 
study (Betzel et al., 2016).
The data processing steps applied to this replication dataset, including the adopted cortical 
parcellation and tapered window, were different from those applied to the HCP dataset.
In this dataset the whole cortex was separated into $114$ regions as in Betzel et al. (2014) and Yeo 
et al. (2015), forming a subdivision of the $17$ network components in Yeo et al. (2011) (see 
Supplementary Fig.~\ref{fig:rois_yeo}).
One region (dorsal prefrontal cortex in the left hemisphere, numbered $84$ in Supplementary 
Fig.~\ref{fig:rois_yeo}) was discarded due to its small surface area and the remaining $113$ regions 
were used as nodes in connectivity analyses.
Time-resolved functional connectivity was computed using an exponential tapered window 
(Zalesky et al., 2014) with a window width of $100$ s, a step size of $1$ TR $=0.645$ s, resulting 
in a total number of $730$ windows.
The setting of tapered window is further detailed in Betzel et al. (2016).

The original data come from Release 1--5 of the NKI-RS\null.
The data were collected with the approval of the institutional review board and all subjects 
provided written informed consent (Nooner et al., 2012).
The original number of subjects in this dataset was $418$ across the life-span.
Using the procedure described in Betzel et al. (2016), a quality controlled sub-sample of healthy 
adults aged $\geq 18$ years and $\leq 30$ years, comprising $80$ participants ($42$ males) was 
extracted.

Imaging data were acquired with a 32-channel head coil on a 3T Siemens Tim Trio scanner.
We utilized rs-fMRI data with the shortest TR of $645$ ms in the NKI dataset 
because fast sampling of rs-fMRI volumes is advantageous when focusing on time-resolved functional 
connectivity.
The rs-fMRI data were collected in a single run of about $10$ min ($900$ time points) in an eyes 
open condition with the following scanning parameters: TE $=30$ ms, flip angle 
$=60^\circ$, voxel size $=3$ mm isotropic, FOV $=222\times222$ mm$\null^2$ and $40$ 
slices.
A T1-weighted structural image was collected with TR $=1{,}900$ ms, TE $=2.52$ ms, flip angle 
$=9^\circ$, voxel size $=1$ mm isotropic, FOV $= 250\times250$ mm$\null^2$ and $176$ slices.

As described in Betzel et al. (2016), the NKI dataset was preprocessed using the Connectome 
Computation System pipeline (CCS; \url{https://github.com/zuoxinian/CCS}) (Xu et al., 2015).
The CCS pipeline incorporates functions included in standard neuroimaging software: AFNI 
(Cox, 2012), Freesurfer (Fischl, 2012), FSL (Jenkinson et al., 2012) and SPM (Ashburner, 2012).
The preprocessing steps of rs-fMRI data included (1) discarding of the first volumes of $10$ s, 
(2) removal of outlier volumes and interpolation, (3) slice timing and motion 
correction, (4) global mean intensity normalization, (5) co-registration between individual 
functional and structural images, (6) nuisance regression using global, white matter and CSF mean 
signals and the Friston-24 motion time series (Friston et al., 1996), (7) temporal band-pass filtering 
($0.01$--$0.1$ Hz), (8) removal of linear and quadratic trends and (9) projection of the 
preprocessed time series onto standard volumetric (MNI152) and cortical surface (fsaverage5) 
spaces.
The low-cut frequency in the temporal filtering ($0.01$ Hz) was specified as the reciprocal of the 
width of the time window ($100$ s).

\section*{Supplementary Results}
\subsection*{Relations to head movements}
\begin{figure}[tb]
  \centering
  \includegraphics[scale=0.62]{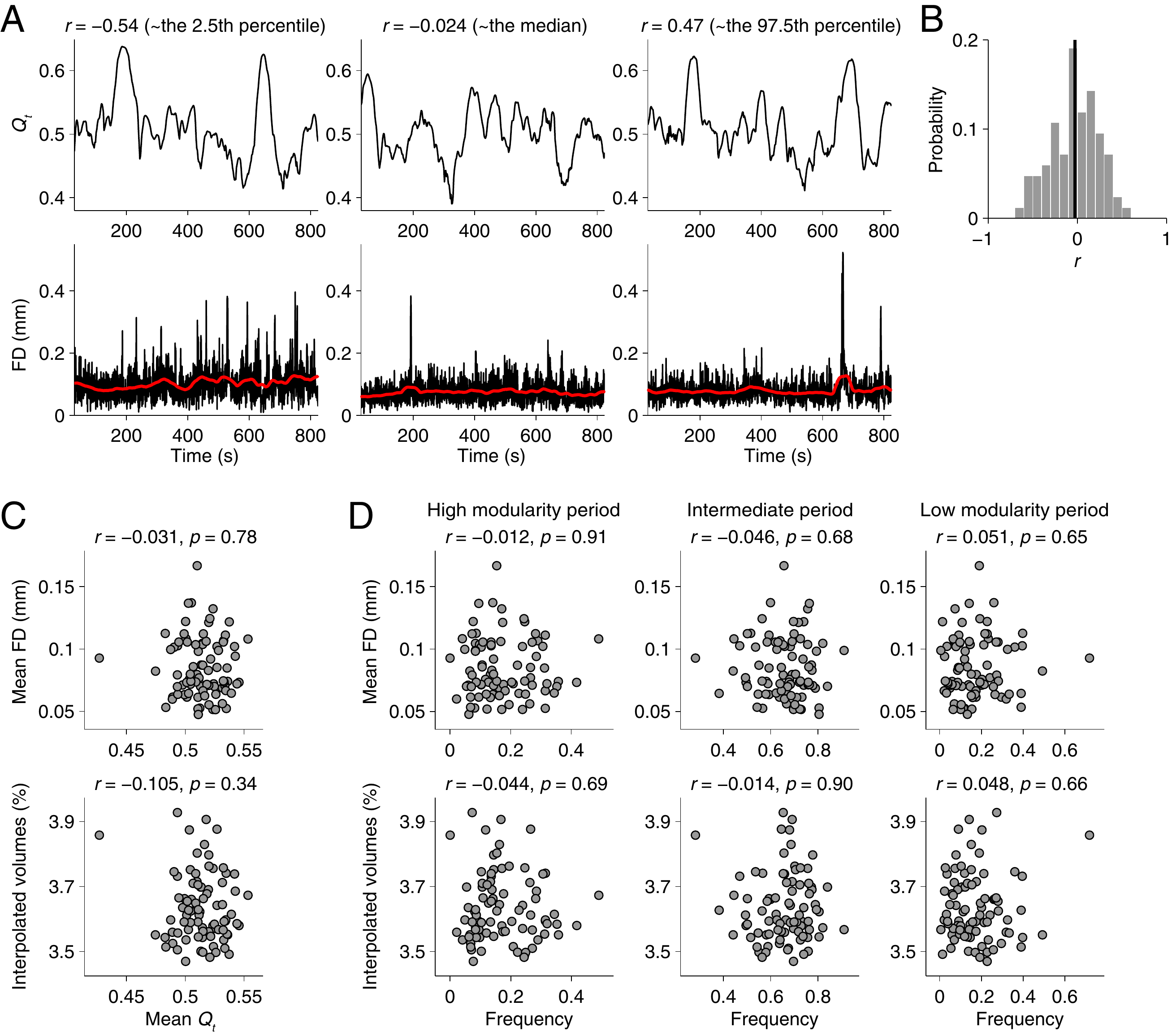}
\caption
 {Head movements and temporal fluctuations in modularity.
 $\vm A$, Upper: time-resolved modularity $Q_t$. Lower: FD and its sliding 
 window average (black and red time courses, respectively).
 We present $Q_t$ and FD in the three subjects whose Pearson correlation coefficient between 
 $Q_t$ and the sliding-window-averaged FD was around the $2.5$th percentile, the median and the 
 $97.5$th percentile of the distribution of this correlation coefficient across subjects.
 $\vm B$, A histogram showing distributions of the correlation coefficient between $Q_t$ and 
 the sliding-window-averaged FD\null.
 The vertical line near zero indicates the mean of the correlation coefficients.
 $\vm C$, Upper: a scatter plot of $Q_t$ and FD, both of which were averaged over time.
 Lower: a scatter plot of mean $Q_t$ and the percentage of interpolated volumes.
 $\vm D$, Upper: scatter plots of mean FD and the frequency of each modularity period.
 Lower: scatter plots of the percentage of interpolated volumes and the frequency of each modularity 
 period.}
\label{fig:motion}
\end{figure}

To demonstrate that the findings in this study are not merely an artifact of head movements, we 
further evaluated the relationships between time-resolved modularity $Q_t$ and FD in run 2LR of the 
HCP dataset.
Supplementary Figure \ref{fig:motion}A shows pairs of the time series of $Q_t$ and FD in the three 
subjects whose Pearson correlation coefficient between $Q_t$ and the sliding-window-averaged FD was 
around the $2.5$th percentile, the median and the $97.5$th percentile of the distribution of this 
correlation coefficient across subjects.
As seen in these time series, peaks and dips in $Q_t$ and FD were not consistently related to each 
other.
Furthermore, the correlation coefficient between $Q_t$ and the sliding-window-averaged 
FD was neither consistently positive nor negative; this correlation coefficient averaged over 
subjects was nearly zero ($-0.029$) (see Supplementary Fig.~\ref{fig:motion}B).
The time-lagged correlation was also very weak (for time lags from $-30$ s to $30$ s, the 
correlation averaged across subjects ranged $-0.052$ to $-0.016$).
Moreover, no clear relationship was found between $Q_t$ and FD when they were averaged over time.
The Pearson correlation coefficient between mean $Q_t$ and mean FD was $-0.031$ 
(see the upper panel in Supplementary Fig.~\ref{fig:motion}C).
In addition, the correlation between mean $Q_t$ and the percentage of interpolated volumes in data 
preprocessing was weak (see the lower panel in Supplementary Fig.~\ref{fig:motion}C).
Mean FD and the percentage of interpolated volumes were not predictive of the frequency of each 
modularity period as well (see Supplementary Fig.~\ref{fig:motion}D).
These observations demonstrate that $Q_t$ (or the frequency of each modularity period) and FD (or 
the percentage of interpolated volumes) in our quality-controlled data were not consistently related, 
indicating that residual head-movement-related noise did not compromise the findings in this study 
about the high and low modularity periods.

\subsection*{Reproducibility of spatial connectivity patterns}
\begin{figure}[tb]
  \centering
  \includegraphics[scale=0.62]{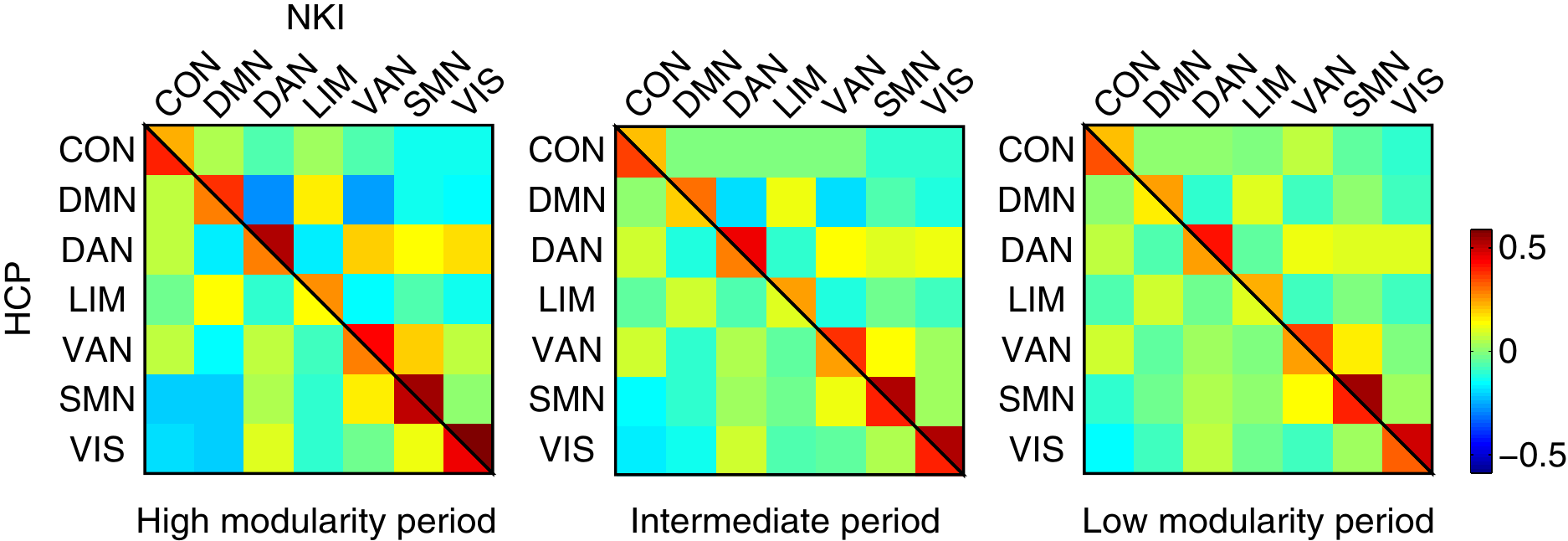}
\caption
 {Edge weights in centroids averaged within each pair of the $7$ network components in Yeo et al. 
 (2011). 
 The lower triangular part is for the HCP dataset (run 2LR) and the higher triangular part is for 
 the NKI dataset.}
\label{fig:qstates_ave}
\end{figure}

We confirmed that spatial connectivity patterns observed in the centroids were reproducible across 
runs and datasets.
In the HCP dataset almost identical patterns were observed in the centroids across all runs 
(see Fig.~\ref{fig:qstates}A and the top three rows of Supplementary Fig.~\ref{fig:qstates_si}).
The clearer dissociation of the DMN from the task-positive networks, DAN, VAN, SMN and VIS, in the 
high modularity period was also observed in the NKI dataset despite the different parcellation used 
(see the bottom row of Supplementary Fig.~\ref{fig:qstates_si}).
To compare the connectivity patterns in different parcellations, the centroid edge weights in the 
HCP and the NKI datasets were averaged within each pair of the $7$ network components in 
Yeo et al. (2011).
Supplementary Figure \ref{fig:qstates_ave} shows that similar module configurations were observed in 
the high modularity period in both datasets, while the decoupling of the DMN from the primary 
sensory networks (SMN and VIS) was more pronounced in the HCP dataset and the decoupling of the 
DMN from the attention networks (DAN and VAN) was more evident in the NKI dataset.
The flat connectivity patterns in the low modularity period were also consistently observed across 
all runs and datasets (see Supplementary Figs.~\ref{fig:qstates_si} and \ref{fig:qstates_ave}).

\subsection*{Reproducibility of temporal homogeneity}
We also confirmed a high reproducibility of the greatest temporal homogeneity of time-resolved 
functional connectivity in the high modularity period.
In all rs-fMRI runs of the HCP and the NKI datasets, the high modularity period exhibited greater 
similarity scores compared to the low modularity period, both for edge weights and partitions (see 
Fig.~\ref{fig:violinplots} and Supplementary Fig.~\ref{fig:violinplots_si}).
The range of the similarity scores shown in Supplementary Fig.~\ref{fig:violinplots_si}A was overall 
larger in the NKI dataset, which may be due to the larger width (HCP, $47.52$ s; NKI, $100$ s) and 
the smaller step duration (HCP, $2.16$ s; NKI, $0.645$ s) of the time window. Nonetheless, the 
relationship of the high and low modularity periods in these similarity scores was invariant across 
the HCP and the NKI datasets.


\newpage
\section*{Supplementary Figures}
\begin{figure}[h]
  \centering
  \includegraphics[scale=0.62]{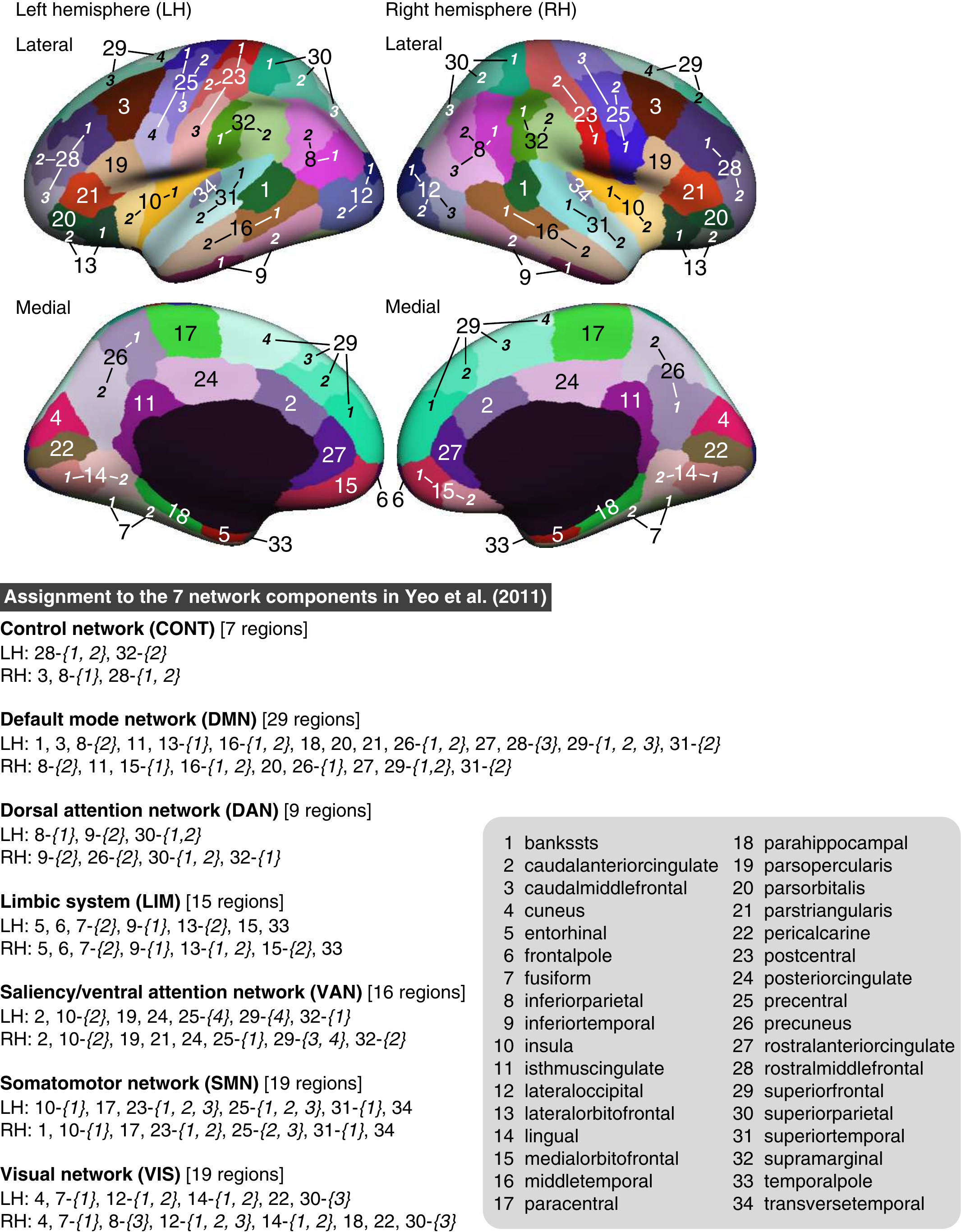}
\caption
 {Cortical parcellation adopted in the HCP dataset, where the $34\times 2$ cortical regions in the 
 Desikan-Killiany atlas in FreeSurfer (listed in the gray panel) are subdivided into a total of 
 $114$ parcels.
 The list below the cortical surface maps shows the assignment of parcels to the maximally 
 overlapped network component in the 7-Network parcellation in Yeo et al. (2011).
 This list is used for sorting regions in connectivity matrices in Figs.\ \ref{fig:qstates} and 
 \ref{fig:similarity_eg} as well as Supplementary Fig.\ \ref{fig:qstates_si}.}
\label{fig:rois_lausanne}
\end{figure}

\newpage
\begin{figure}[tb]
  \centering
  \includegraphics[scale=0.62]{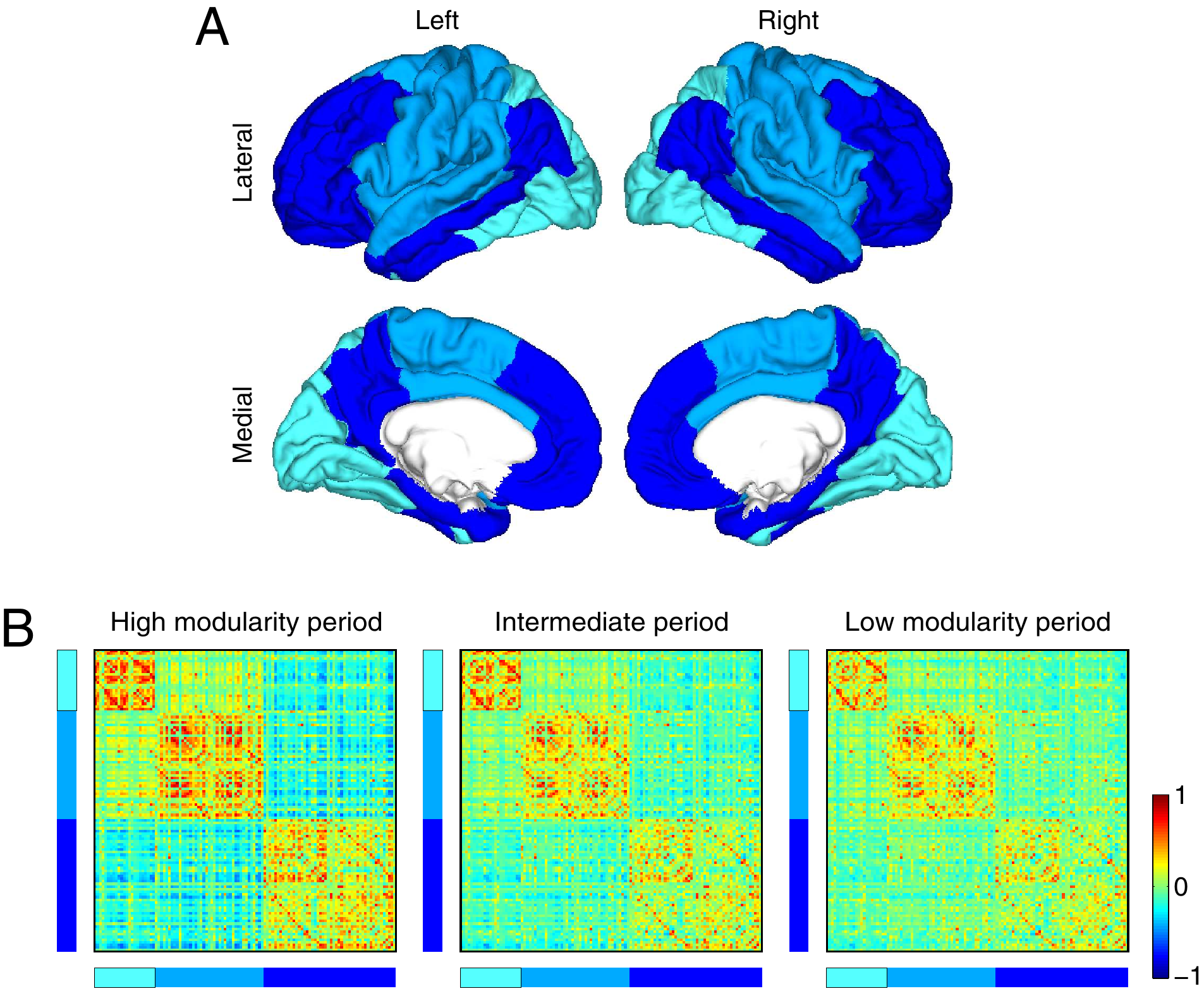}
\caption
 {$\vm A$, Communities detected by modularity maximization (with the default resolution 
 parameter $\gamma=1$) from the long-timescale functional connectivity averaged over all subjects.
 The cyan module contained visual areas (task-positive), the light blue module contained motor, 
 somatosensory and auditory areas (task-positive), and the blue module contained regions in the 
 default mode network (task-negative).
 $\vm B$, Centroids of time-resolved functional connectivity presented in Fig.~\ref{fig:qstates}A, 
 where nodes in each connectivity matrix are reordered according to the partition shown above.}
\label{fig:reordered_centroids}
\end{figure}
\null

\newpage
\begin{figure}[tb]
  \centering
  \includegraphics[scale=0.62]{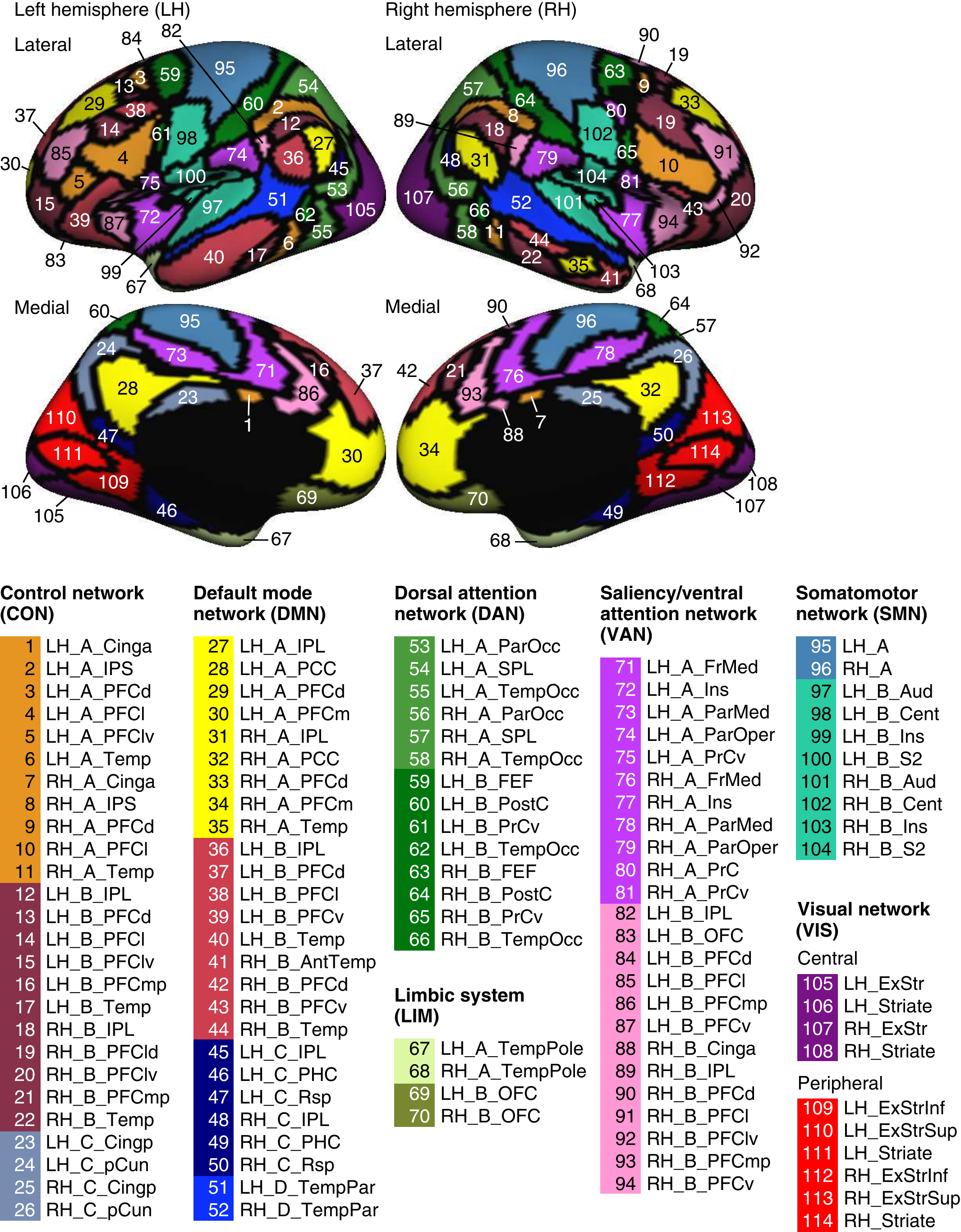}
\caption
 {Cortical parcellation adopted in the NKI dataset, where the 17 network components in Yeo et al. 
 (2011) are subdivided into $114$ regions.
 The list of abbreviations for cortical regions is presented in Supplementary Table 
 \ref{tab:abbreviations}.
 A parcel numbered $84$ was discarded in this study because of its small surface area.
 Regions in connectivity matrices in Supplementary Fig.\ \ref{fig:qstates_si} are sorted according 
 to the numbering of parcels in this figure.}
\label{fig:rois_yeo}
\end{figure}
\null

\newpage
\null

\newpage
\begin{figure}[tb]
  \centering
  \includegraphics[scale=0.62]{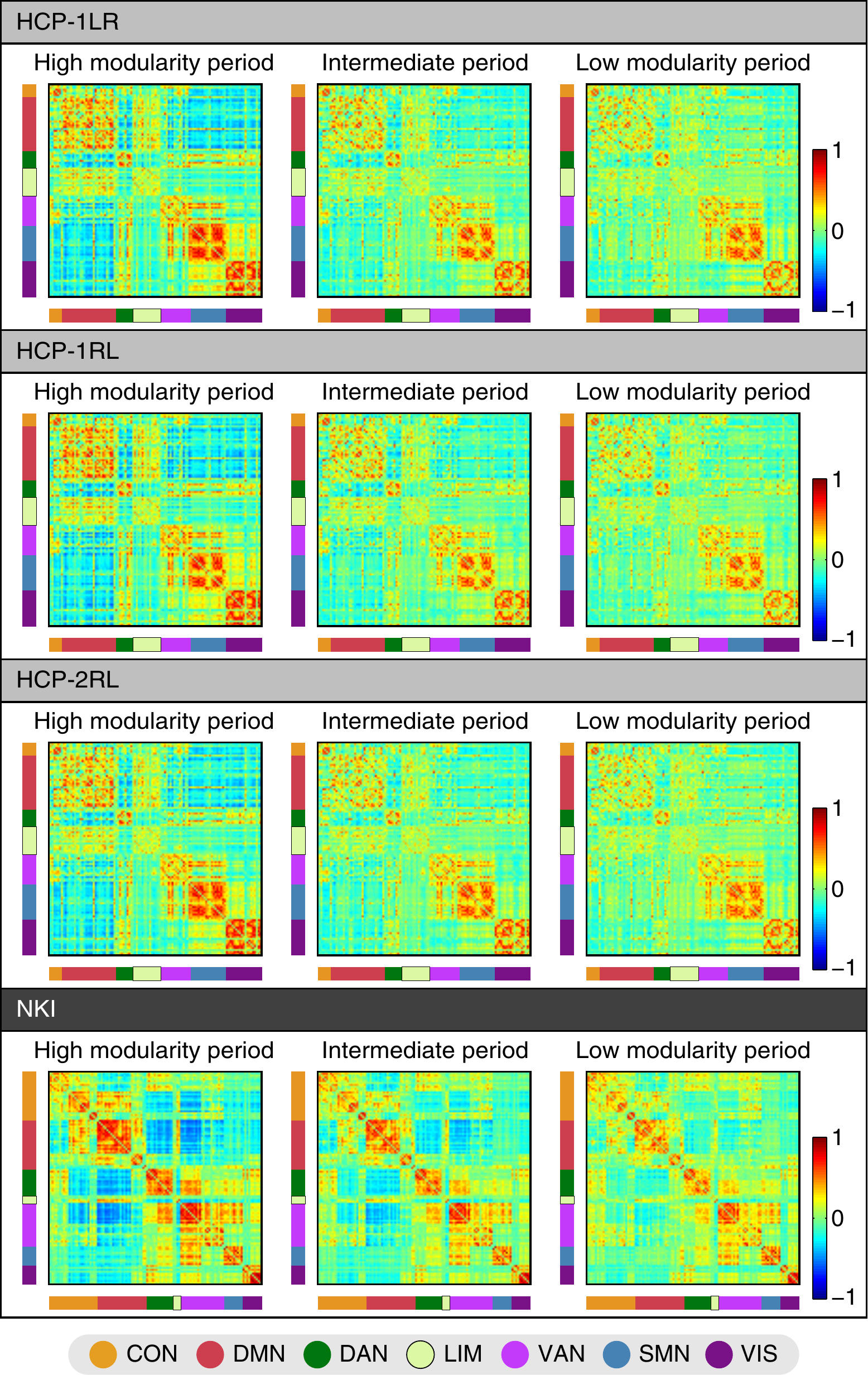}
\caption
 {Centroids of time-resolved functional connectivity in the high modularity, the intermediate and 
 the low modularity periods, in the HCP and the NKI datasets.
 Results in run 2LR of the HCP dataset are shown in Fig.~\ref{fig:qstates}A.
 Note that the cortical parcellations adopted in the HCP and the NKI datasets were different from 
 each other.}
\label{fig:qstates_si}
\end{figure}
\null

\newpage
\null

\newpage
\begin{figure}[tb]
  \centering
  \includegraphics[scale=0.62,angle=90]{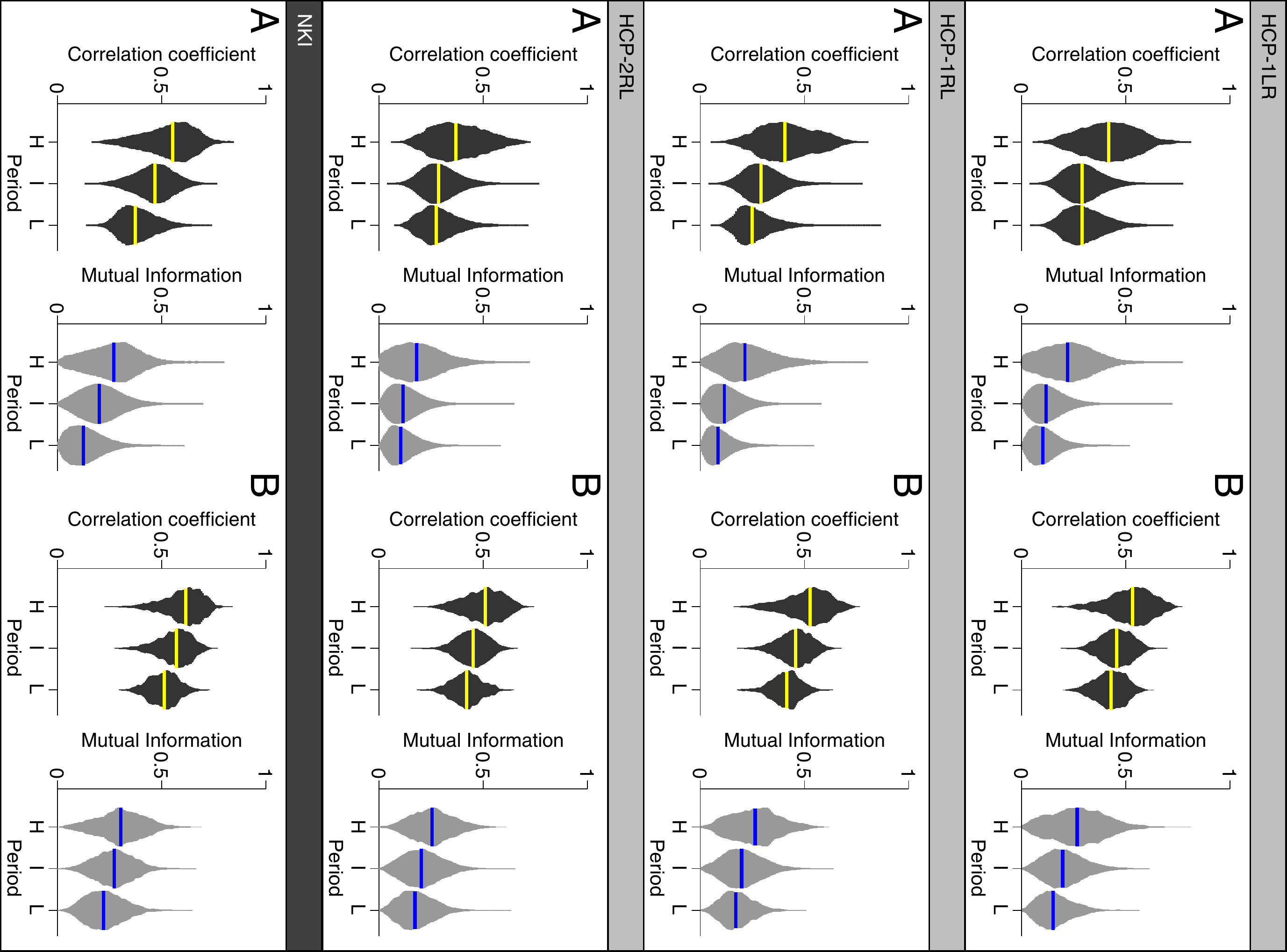}
\caption
 {Distributions of the similarity measures of time resolved functional connectivity in the HCP and 
 the NKI datasets.
 Results in run 2LR of the HCP dataset are shown in Fig.~\ref{fig:violinplots}.
 A bar in a distribution indicates the median.
 {$\vm A$}, The similarity of edge weights (left) and partitions (right) between samples of 
 time-resolved functional connectivity.
 These distributions were derived only from pairs of samples that are apart from each other by more 
 than the width of the time window.
 {$\vm B$}, The similarity of edge weights (left) and partitions (right) between each sample of 
 time-resolved functional connectivity and its corresponding centroid.}
\label{fig:violinplots_si}
\end{figure}

\newpage
\section*{Supplementary Tables}
\renewcommand{\arraystretch}{0.8}
\begin{table}[h]
 \centering
\caption
 {Abbreviations in Supplementary Fig.\ \ref{fig:rois_yeo}.}\vspace{0mm}
\label{tab:abbreviations}
\small
 \begin{tabular}{ll}
  \toprule
   Abbreviation \ \ \ \ \ \ \ \ \ \ \ \ \ \ \ \ \ \ \ \ \ \ \ \ \ \ \ \ \ \ \ \ \ \ \ & Region name\\
  \midrule
AntTemp	 &	Anterior temporal cortex\\         
Aud	 &	Auditory cortex\\                 
Cent	 &	Central sulcus\\                 
Cinga	 &	Anterior cingulate cortex\\     
Cingp	 &	Posterior cingulate cortex\\      
ExStr	 &	Extrastriate cortex\\               
ExStrInf &	Inferior extrastriate cortex\\      
ExStrSup &	Superior extrastriate cortex\\      
FEF	 &	Frontal eye fields\\               
FrMed	 &	Medial frontal cortex\\           
Ins	 &	Insula\\                         
IPL	 &	Inferior parietal lobule\\         
IPS	 &	Intraparietal sulcus\\            
OFC	 &	Orbitofrontal cortex\\             
ParMed	 &	Medial parietal cortex\\         
ParOcc	 &	Parieto-occipital cortex\\         
ParOper	 &	Parietal operculum\\               
PCC	 &	Posterior cingulate cortex\\      
pCun	 &	Precuneus\\                        
PFCd	 &	Dorsal prefrontal cortex\\          
PFCl	 &	Lateral prefrontal cortex\\         
PFCld	 &	Dorsolateral prefrontal cortex\\    
PFClv	 &	Ventrolateral prefrontal cortex\\   
PFCm	 &	Medial prefrontal cortex\\          
PFCmp	 &	Posterior-medial prefrontal cortex\\
PFCv	 &	Ventral prefrontal cortex\\       
PHC	 &	Parahippocampal cortex\\          
PostC	 &	Post-central cortex\\             
PrC	 &	Pre-central cortex\\               
PrCv	 &	Ventral pre-central cortex\\       
Rsp	 &	Retrosplenial cortex\\             
S2	 &	Secondary somatosensory cortex\\   
SPL	 &	Superior parietal lobule\\        
Striate	 &	Striate cortex\\                  
Temp	 &	Temporal cortex\\                   
TempOcc	 &	Temporo-occipital cortex\\         
TempPar	 &	Temporo-parietal cortex\\           
TempPole &	Temporal pole\\
  \bottomrule
 \end{tabular}
\end{table}
\renewcommand{\arraystretch}{1}
\null
\end{document}